\documentclass[12pt]{article}
\usepackage[dvipdfmx]{graphicx}
\usepackage{amsmath,amsthm,amsfonts,amssymb}
\usepackage{color}
\usepackage[normalem]{ulem}
\usepackage{arydshln, mathrsfs}
\usepackage{wrapfig}
\usepackage{dutchcal}
\usepackage{subfigure}

\theoremstyle{definition}

\newtheorem*{theorem*}{Theorem}

\newtheorem*{definition*}{Definition}

\makeatletter
\@addtoreset{equation}{section}

\makeatother

\usepackage{appendix}
\usepackage{tikz}
\usepackage{soul}

\usepackage{color}

\begin{document}
	
	\begin{titlepage}
		\null
		\begin{flushright}
			arXiv:2501.08250
			\\
			April, 2025
		\end{flushright}
		
		\vskip 0.6cm
		\begin{center}
			
			{\LARGE \bf Soliton  Resonances in Four Dimensional \\
				
				\vskip 0.5cm
				
				Wess-Zumino-Witten Model}

			\vskip 1cm
			\normalsize
			
			{\large 
				Shangshuai Li\footnote{E-mail:lishangshuai@shu.edu.cn}{}$^{,a,b}$,
				Masashi Hamanaka\footnote{E-mail:hamanaka@math.nagoya-u.ac.jp}{}$^{,c}$,\\\vspace{2mm}
				Shan-Chi Huang\footnote{E-mail:x18003x@math.nagoya-u.ac.jp (Co-first author)}{}$^{,c}$,
				Da-Jun Zhang\footnote{E-mail:djzhang@staff.shu.edu.cn}{}$^{,a,d}$.
				
			}
			
			\vskip 0.7cm
			
			{\it  
				{}$^{a}$Department of Mathematics, Shanghai University, Shanghai, 200444, CHINA, \\
\vspace{2mm}
				{}$^{b}$Department of Applied Mathematics, Faculty of Science and Engineering, \\
				Waseda University, Tokyo, 169-8555, JAPAN, \\
\vspace{2mm}
				{}$^c$Graduate School of Mathematics, Nagoya University,
				Nagoya, 464-8602, JAPAN, \\
\vspace{2mm}
				{}$^d$ Newtouch Center for Mathematics of Shanghai University,  Shanghai, 200444, CHINA
				}
			
			
			
			\vskip 0.7cm
			
			{\bf \large Abstract}
			\vskip 0.3cm
			
		\end{center}

We present two kinds of resonance soliton solutions on the Ultrahyperbolic space $\mathbb{U}$ for the $G=U(2)$ Yang equation, which is equivalent to the anti-self-dual Yang-Mills (ASDYM) equation. 
We  reveal and  illustrate the solitonic behaviors in the four-dimensional Wess-Zumino-Witten (WZW$_4$) model through the sigma model action densities. 
 The Yang equation is the equation of motion of the WZW$_4$ model.  
In the case of $\mathbb{U}$, the WZW$_4$ model   describes a string field theory action of open $N=2$ string theories. Hence, our solutions  on $\mathbb{U}$   suggest the existence of the corresponding classical objects in the $N=2$ string theories. Our solutions include multiple-pole solutions and V-shape soliton solutions. The V-shape solitons suggest annihilation and creation processes of two solitons and would be building blocks to classify the ASDYM solitons, like the role of Y-shape solitons in classification of the KP (line) solitons.

We also clarify the relationship between the Cauchy matrix approach and the binary Darboux transformation in terms of quasideterminants. Our formalism can start with a simpler input data for the soliton solutions and hence might give a suitable framework for the classification of the ASDYM solitons. 

	\end{titlepage}
	
	\clearpage
	\baselineskip 6.1mm
	
	
	\newpage
				
	

%
%
\tableofcontents

\section{Introduction}

Anti-self-dual Yang-Mills (ASDYM) equations have been attracting great interest for many years and established central positions in mathematics and physics. In gauge theories, instantons, global solutions to the ASDYM equation, have played crucial roles in revealing non-perturbative aspects of them \cite{Shifman-1994}  and have brought a new perspective to four-dimensional geometry (e.g., \cite{DK-1990}). In integrable systems, the ASDYM equation is a master equation in the sense that it can be reduced to various lower-dimensional integrable equations, known for the Ward's conjecture \cite{Ward-1985}. Reduced equations include the Calogero-Bogoyavlenski-Shiff (CBS) equation and the Zakharov system in $(2+1)$ dimensions, the Korteweg-de Vries (KdV) equation, the non-linear Schr\"odinger (NLS) equation and Toda equations in $(1+1)$ dimensions, and Painlev\'e equations in $(0+1)$ dimensions. (see \cite{Mason-1996} and references therein). 
One recent example of the reduction is the Fokas-Lenells equation \cite{LLZ-2024}, which is related to the Kaup-Newell spectral problem and the derivative nonlinear Schr\"{o}dinger equation family \cite{Lenells-2009-1,Lenells-2009-2}.
The ASDYM equation itself is integrable in various senses (e.g., \cite{Ward-1977})--
\cite{Sasa-1998}. 
Furthermore, there is another attractive feature of the ASDYM equation in string theories. The ASDYM equation is equivalent to the Yang equation \cite{Yang-1977, BFNY-1978} which is the equation of motion of the four-dimensional Wess-Zumino-Witten (WZW$_4$) model \cite{Donaldson-1985}--
\cite{Inami-1997-T}. In the case of the split signature $(+,+,-,-)$, the WZW$_4$ model describes the space-time action of the open $N=2$ string theory \cite{Ooguri-1991-H}--
\cite{Berkovits-1996} and, therefore, solutions of the ASDYM can be realized and applicable for this string theory as classical physical objects.

Two of the authors have recently constructed soliton solutions of the ASDYM equations (based on the result of Darboux transformations \cite{Nimmo-2000, Gilson-2020}) and calculated the action densities of them in order to clarify the solitonic behaviors 
\cite{HH-2020}--
\cite{Hamanaka-2024}. It is proven that the soliton solutions of them are described by Wronskian-type quasideterminants (called quasi-Wronskians) \cite{Gelfand-1991} and the ASDYM solitons behave quite similar to the KP (line) solitons in the sense of real-valued action densities: the one-soliton solutions are codimension-one solitons, whose action densities are localized on three-dimensional hyperplanes in four dimensions. The $n$-soliton solutions can be interpreted as a ``non-linear superposition'' of $n$ one-soliton solutions with phase shifts. These ASDYM solitons are considered as non-resonance solutions, called ``line solitons'' in this paper. Real-valued KP solitons are classified in terms of positive Grassmannians \cite{Kodama-Williams-2014, Kodama-2017} and applied to shallow water waves \cite{Kodama-2018}. We can hence expect that similar classification of ASDYM solitons would be possible and could be applied to reveal non-perturbative aspects of the open $N=2$ string theory by analyzing the moduli spaces of them. 

On the other hand, another two of the authors have successfully applied the Cauchy matrix approach (CMA) to the ASDYM equations to construct Grammian-type solutions \cite{LQYZ-SAPM-2022, LSS-2023}. In the CMA, the input data are simpler than those in the Darboux transformation and would be more suitable for the classification of the ASDYM solitons. For the (scalar) KP equation, it is proved that Wronskian and Grammian soliton solutions are equivalent \cite{CLM-2010} and it is worth clarifying whether this is true of the ASDYM solitons or not. The CMA closely relates to the binary Darboux transformation \cite{Nimmo-2000}, direct linearization methods \cite{LSS-2024} and the bi-differential calculus \cite{FMH-2024}. It is also worthwhile to rewrite their formalisms in terms of quasideterminants so as to clarify the relationship between the CMA and the binary Darboux transformation because quasideterminants usually simplify the calculations and make it easier to get to the essence.  

In this paper, we construct resonance solutions in the context of the WZW$_4$ model. 
In integrable systems, resonance solutions can usually be realized as limits of multi-line solitons, such as large phase-shift limits. 
In fact, phase shifts come from ``intermediate states,'' as discussed in Sec. \ref{sec-4-1}. Hence, we can expect that in the large phase-shift limit of the two-soliton, 
the intermediate state can be observed as a one-soliton.
This expectation is actually correct for complex valued solutions, however by putting the reality condition on the solutions, the amplitudes of the intermediate states become zero. This 
fact implies that in the large phase-shift limit, our two-soliton solutions decompose into two V-shape solitons, not the composite of two Y-shape solitons. 
Our result is consistent with the fact that there actually exist V-shape solitons in the CBS equation \cite{FSTY-1998} and the Zakharov system \cite{Strachan-1992-W}, both of which can be obtained from the ASDYM equation by dimensional reduction while, the KP equation cannot. Existence of V-shape solitons of the ASDYM equation suggests a quite different aspect from the KP solitons, and classification of the ASDYM solitons might be achieved based on non-Grassmannian geometry. 
On the other hand, by tuning orientations of the two-line solitons, we get another type of resonance solution, the double-pole solution \cite{Wadati-1982}, which describes the interaction between two solitary waves of equal amplitude \cite{Gagnon-1992}. By observing the limit processes, we find  suitable input data for the multiple-pole solution, where the spectral parameter matrix is in a Jordan normal form. 

In order to  make interpretations of the solutions in the WZW$_4$ model, we restrict our discussion to the case of the split signature 
 and of unitary group valued solutions (simply called unitary solutions). 
The WZW$_4$ model consists of two terms: the non-linear sigma model (NL$\sigma$M) term and the Wess-Zumino term. 
For the ASDYM solitons, 
the main contribution of solitonic behavior can be captured by the NL$\sigma$M term because the Wess-Zumino term identically vanishes in the asymptotic region. 
Furthermore, in a dimensionally reduced system, the Hamiltonian of the Wess-Zumino term always vanishes everywhere \cite{Hamanaka-2023}. 
Therefore, calculating the NL$\sigma$M term is  sufficient for the illustration of solitonic behavior.
On the other hand,
the unitary solutions lead to real-valued nonsingular NL$\sigma$M terms and, hence, the action densities can be plotted specifically by \textit{Mathematica}. 
We also clarify the relationship between the CMA and the binary Darboux transformation in terms of quasideterminants. Our formalism can start with simpler input data for the soliton solutions and, hence, might give a suitable framework for the classification of the ASDYM solitons. 

This paper is organized as follows. In Sec. 2, we give a brief review of the WZW$_4$ model. In Sec. 3, we reformulate the CMA in terms of quasideterminants. We extend the dispersion relation in the Cauchy matrix approach so that the relation to the binary Darboux transformation is clarified in  Sec. 3.4. Finally we get unitary quasi-Grammian solutions. 
In  Sec. 4. we present exact one- and two-soliton solutions  by the formulation of quasi-Grammian. The new input data are much simpler than those used in the Darboux transformations, which are formulated by quasi-Wronskian. 
We calculate the NL$\sigma$M action density for one- and  two-solitons respectively, and find that our results coincide with the ones calculated from quasi-Wronskian solutions \cite{Hamanaka-2023}.
In  Sec. 4.3, we discuss some particular limits of the solutions which lead to resonance processes.  In  Sec. 5, we construct the multiple-pole solution from the modified input data  where the spectral parameter matrix is in a Jordan normal form.   
In  Sec. 6, 
the distributions of action densities for various resonance solutions are demonstrated as 2D slice figures by using \textit{Mathematica}.
These figures show that the double-pole solutions actually have non-trivial behavior of resonances or bound states and V-shape solitons emerge in the large  phase-shift limits of the two-soliton.
 Sec. 7 is devoted to the conclusion and discussion. The Appendix includes some miscellaneous formulas, detailed calculations, and necessary proofs.


\section{Four-Dimensional Wess-Zumino-Witten Model}

In this section, we present a brief review of the WZW$_4$ model. For more details, refer to \cite{Hamanaka-2023}.
For convenience of discussion, firstly, we introduce a four-dimensional space $\mathbb{M}_4$ with complex coordinates $(z,\widetilde z, w, \widetilde w)$ and the flat metric \cite{Mason-1996}, defined by
\begin{eqnarray}
	ds^2:=g_{mn}dz^{m}dz^{n}:=2(dz d\widetilde z -dw d\widetilde w),
	\end{eqnarray}
where $m,n\in \{1,2,3,4\}$ and $(z^{1},z^{2},z^{3},z^{4}):=(z,\widetilde z, w, \widetilde w)$. 

For our purpose in this paper, we focus our discussion only on the four-dimensional flat real space with the split signature, which is called the Ultrahyperbolic space denoted by $\mathbb{U}$. With the local coordinates $(x^1,x^2,x^3,x^4)$ of $\mathbb{U}$, the metric can be chosen as follows: 
\begin{eqnarray}
\label{metric_U}
ds^2=\eta_{\mu\nu}dx^{\mu}dx^{\nu}:=(dx^{1})^2+(dx^{2})^2-(dx^{3})^2-(dx^{4})^2,
\end{eqnarray}
where $\mu,\nu\in \{1,2,3,4\}$. 
This real space $\mathbb{U}$ can be realized from $\mathbb{M}_4$ 
by imposing the following reality condition on $(z,\widetilde z, w, \widetilde w)$:
\begin{eqnarray}
\label{Reality condition_U}
&(\mathbb{U})&
\left(
\begin{array}{cc}
z & w \\
\widetilde{w} & \widetilde{z}
\end{array}\right)
=
\displaystyle{\frac{1}{\sqrt{2}}}
\left(\begin{array}{cc}
x^{1}+x^{3} & x^{2}+x^{4} \\
-(x^{2}-x^{4}) & x^{1}-x^{3} 
\end{array}\right), \label{U}
\end{eqnarray}
where $z, \widetilde{z}, w, \widetilde{w} \in \mathbb{R}$.\footnote{
We note that there is another real slice $\widetilde z=\overline z$ and $\widetilde w= \overline w$ to realize the split signature, however,  
we do not consider this case in this paper because the unitarity condition of $J$ leads to trivial action densities \cite{Hamanaka-2023}.}
If we replace the reality condition \eqref{Reality condition_U} with 
$\widetilde z=\overline z$ and $\widetilde w= -\overline w$, 
we can discuss the Euclidean case \cite{Hamanaka-2023}. 

Now, we consider the action of the WZW$_4$ model defined on  $\mathbb{U}$,   denoted by $S_{\scriptsize{\mbox{WZW}_4}}$. The WZW$_4$ action consists of two parts, that is, the NL$\sigma$M term $S_{\sigma}$ and the Wess-Zumino term $S_{\scriptsize{\mbox{WZ}}}$, which are explicitly given by
\begin{eqnarray}
S_{\scriptsize{\mbox{WZW}_4}}
&\!\!\!\!:=\!\!\!\!&S_{\sigma} + S_{\scriptsize{\mbox{WZ}}} \nonumber
\\
&\!\!\!\!:=\!\!\!\!&
\frac{i}{4\pi}\int_{ \mathbb{U} }\omega \wedge \mbox{Tr}\left[
(\partial J)J^{-1} \wedge (\widetilde{\partial}J)J^{-1}
\right]
-
\frac{i}{12\pi}\int_{ \mathbb{U}} A \wedge \mbox{Tr}\left[(dJ)J^{-1}
\right]^3,
\end{eqnarray} 
where the dynamical variable $J$ is a smooth map from  $\mathbb{U}$   
 to $G=GL(N, \mathbb{C})$, and $\omega$ is 
a two-form on $\mathbb{U}$ given by 
\begin{eqnarray}
\omega = \frac{i}{2}(dz \wedge d\widetilde{z} - dw \wedge d\widetilde{w}).
\end{eqnarray} 
The differential one-form $A$ is chosen as 
$A=(i/4)(zd\widetilde{z}-\widetilde{z}dz - wd\widetilde{w} +\widetilde{w}dw)$ so that $\omega=dA$. 
The cubic term $[(dJ)J^{-1}]^3$ is an abbreviation of the wedge product of three  one-form: 
$\left[(dJ)J^{-1}\right]^3=(dJ)J^{-1} \wedge (dJ)J^{-1} \wedge (dJ)J^{-1}$. 
The exterior derivatives are defined as
\begin{eqnarray}
d:=\partial + \widetilde{\partial}, ~ 
\partial:=dw\partial_{w} + dz\partial_{z},~
\widetilde{\partial}:=d\widetilde{w}\partial_{\widetilde{w}} + d\widetilde{z}\partial_{\widetilde{z}}.
\end{eqnarray}
The equation of motion of the WZW$_4$ model is exactly the Yang equation,
\begin{eqnarray}
\label{EOM}
\widetilde{\partial}\left[
\omega \wedge (\partial J)J^{-1}
\right]=0
~\Longleftrightarrow ~
\partial_{\widetilde{z}}\left[
(\partial_{z}J)J^{-1}
\right]
-
\partial_{\widetilde{w}}\left[
(\partial_{w}J)J^{-1}
\right]
=0,
\end{eqnarray}
which gives an equivalent expression for the anti-self-dual Yang-Mills equation.

The WZW$_4$ action can be expressed in terms of the real coordinates. 
For the NL$\sigma$M action, we have
\begin{eqnarray}
S_{\sigma}
&\!\!\!\!=\!\!\!\!&
\frac{-1}{16\pi}\int_{\mathbb{U}}\mbox{Tr}\left[
(\partial_{m}J)J^{-1}(\partial^{m}J)J^{-1}
\right]dz \wedge d\widetilde{z} \wedge dw \wedge d\widetilde{w} \nonumber
\\
&\!\!\!\!=\!\!\!\!&
\frac{-1}{16\pi}\int_{\mathbb{U}}\mbox{Tr}\left[
(\partial_{\mu}J)J^{-1}(\partial^{\mu}J)J^{-1}
\right]dx^{1} \wedge dx^{2} \wedge dx^{3} \wedge dx^{4},
\end{eqnarray} 
where $\partial^{m}:=g^{mn}\partial_{n}$ and $\partial^{\mu}:=\eta^{\mu\nu}\partial_{\nu}$,
in which $g^{mn}$ and  $\eta^{\mu\nu}$  are the inverse matrices of 
$g_{mn}$ and  $\eta_{\mu\nu}$   respectively. For the component representation of the Wess-Zumino action, one can refer to \cite{Hamanaka-2023}.

As mentioned in  Sec. 1, we will only calculate the NL$\sigma$M action density because the main contribution can be captured by this term well.
The NL$\sigma$M action density is real-valued when $J$ is unitary  
because $(\partial_{\mu}J)J^{-1}$ is anti-Hermitian with pure imaginary eigenvalues in this case. 

\section{Quasi-Grammian Solutions in the WZW$_4$ Model}
In this section, we briefly review the CMA for the Yang equation \cite{LQYZ-SAPM-2022,LSS-2023}, and then extend the dispersion relations to a general version. 
For the Ultrahyperbolic space $\mathbb{U}$, we show that such dispersion relations can be regarded as initial linear systems in binary Darboux transformations. Finally, we reduce two linear systems to a single one to construct unitary solutions. 

\subsection{A Brief Review of the Cauchy Matrix Approach ($G=GL(2, \mathbb{C})$)}
The CMA for the Yang equation was developed by two of the authors and collaborators in recent research \cite{LQYZ-SAPM-2022,LSS-2023}.  
This method starts by introducing the Sylvester equation
\begin{eqnarray}
\label{Sylvester eq}
KM(r, s)-M(r,s)L=rs^{T},
\end{eqnarray}
where $K$ and $L$ are constant matrices and 
$r,s$ satisfies the following dispersion relations:
\begin{eqnarray}
\label{Dispersion relations}
\partial_{x_j}r=K^{j}r\sigma_{3}, ~~\partial_{x_j}s=-(L^{T})^{j}s\sigma_{3}, ~~j \in \mathbb{Z}.
\end{eqnarray}
Solutions $M(r,s)$ of the Sylvester equation are dressed Cauchy matrices.\footnote{The Cauchy matrix is a matrix in the form of $\left(\frac{1}{k_i-l_j}\right)_{1\leq i,j\leq N}$ , while the dressed Cauchy matrix is in the form of $\left(\frac{\rho(k_i)\sigma(l_j)}{k_i-l_j}\right)_{1\leq i,j\leq N}$. } 
The sizes of these involved matrices are given so that the above matrix multiplications are well defined. Here, we consider the case of $K, L \in \mathbb{C}_{N \times N}$, $r, s \in \mathbb{C}_{N \times 2}[x], M(r,s) \in \mathbb{C}_{N \times N}[x]$, $\sigma_{3}:=\mbox{diag}(1, -1)$ and $x:=(\cdots, x_{-1}, x_{0}, x_{1}, \cdots)$. The number $N$ relates to the number of solitons as is discussed in  Sec. 4.


Defining  $u,v \in \mathbb{C}_{2\times 2}[x]$  as 
\begin{eqnarray}
\label{u, v}
u=-s^{T}M(r,s)^{-1}r, ~v=I - s^{T}M(r,s)^{-1}K^{-1}r,
\end{eqnarray}
we can prove a differential recurrence relation (see Appendix A in \cite{LSS-2023}),
\begin{eqnarray}
(\partial_{x_{j+1}}v)v^{-1}=\partial_{x_{j}}u,~~ j \in \mathbb{Z}, 
\end{eqnarray}
and therefore, for any given $n, m \in \mathbb{Z}$, through the compatibility 
$\partial_{x_{m}}(\partial_{x_{n}}u)=\partial_{x_{n}}(\partial_{x_{m}}u)$, 
$v$ satisfies
\begin{eqnarray}
\label{Yang eq_mn}
\partial_{x_{m}}[(\partial_{x_{n+1}}v)v^{-1}]
-
\partial_{x_{n}}[(\partial_{x_{m+1}}v)v^{-1}]=0, 
\end{eqnarray}
which is the Yang equation if we assign $(x_{n+1}, x_{m}, x_{m+1}, x_{n})$ to $(z, \widetilde{z}, w, \widetilde{w}) 
$.

In fact, $u$ and $v$ possess the structure of quasi-Grammians (Grammian-like quasideterminants \cite{Gilson-2007}).
To explain this, firstly, we give several equivalent definitions of quasideterminants for 
 the following $(N+k)\times(N+k)$  matrix 
 with partition:
\begin{eqnarray}
\label{Quasideteminant_defn_1}
\left|
\begin{array}{cc}
A_{N \times N} & B_{N \times k} \\
C_{k \times N} & \fbox{$d_{k \times k}$}
\end{array}
\right|
:= 
d - CA^{-1}B  \in \mathbb{C}_{k \times k}, 
\end{eqnarray}
for any positive integers $k,N$. In particular , $k=2$ implies 
\begin{eqnarray}
\label{Quasideteminant_defn_2}
&&\!\!\!\!\left|
\begin{array}{cc}
A & \begin{array}{cc} B_1 & B_2\end{array}
\\
\begin{array}{c} C_1 \\ C_2\end{array} &
\fbox{$
\begin{array}{cc}
d_{11} & d_{12} \\
d_{21} & d_{22}
\end{array}
$}
\end{array}
\right|
\nonumber \\
&\!\!\!\!=\!\!\!\!&
\left(
\begin{array}{cc}
\left|
\begin{array}{cc}
A & B_1 \\
C_1 & \fbox{$d_{11}$}
\end{array}
\right|
&
\left|
\begin{array}{cc}
A & B_2 \\
C_1 & \fbox{$d_{12}$}
\end{array}
\right|
\medskip \\
\left|
\begin{array}{cc}
A & B_1 \\
C_2 & \fbox{$d_{21}$}
\end{array}
\right|
&
\left|
\begin{array}{cc}
A & B_2 \\
C_2 & \fbox{$d_{22}$}
\end{array}
\right|
\end{array}
\right)
=
\frac{1}{|A|}
\left(
\begin{array}{cc}
\left|
\begin{array}{cc}
A & B_1 \\
C_1 & d_{11}
\end{array}
\right|
&
\left|
\begin{array}{cc}
A & B_2 \\
C_1 & d_{12}
\end{array}
\right|
\medskip \\
\left|
\begin{array}{cc}
A & B_1 \\
C_2 & d_{21}
\end{array}
\right|
&
\left|
\begin{array}{cc}
A & B_2 \\
C_2 & d_{22}
\end{array}
\right|
\end{array}
\right),
\end{eqnarray}
where
\begin{align}
	d=d_{2\times2}:=
	\begin{pmatrix}
		d_{11} & d_{12} \\
		d_{21} & d_{22}
	\end{pmatrix},~
	B=B_{N\times2}:=(B_1,B_2),~
	C=C_{2\times N}:=
	\begin{pmatrix}
		C_1 \\
		C_2
	\end{pmatrix}.
\end{align}
 The last equality in \eqref{Quasideteminant_defn_2} is due to 
the commutative limit of quasideterminants (see proposition 2.2 and equality (2.12) in \cite{Huang-2021}).

Now \eqref{u, v} can be rewritten in the form of quasideterminants as
\begin{eqnarray}
\label{u,v_quasi-Grammian}
u=\left|
\begin{array}{cc}
M(r,s) & r \\
s^{T} & \fbox{$0$}
\end{array}
\right|,~~
v=\left|
\begin{array}{cc}
M(r,s) & K^{-1}r \\
s^{T} & \fbox{$I$}
\end{array}
\right|.
\end{eqnarray}
Note that the dressed Cauchy matrix $M(r,s)$ has a Grammian-like structure \cite{Gilson-2007} and can be represented as a difference of two Gram matrices, as follows. Let us represent each column of $r,s$ explicitly as 
\begin{align}	
r:=(r_1,r_2),~~s:=(s_1,s_2),~~r_1,r_2,s_1,s_2\in\mathbb C_{N\times1}[x].
\end{align}
From \cite{LQYZ-SAPM-2022,LSS-2023}, the derivative of $M(r,s)$ can be rewritten as

\begin{align}
\partial_{x_n}M(r,s)&=\displaystyle{\sum_{\ell=0}^{n-1}}K^{n-\ell-1}r\sigma_{3}s^TL^{\ell}=\displaystyle{\sum_{\ell=0}^{n-1}}K^{n-\ell-1}(r_1,r_2)\mathrm{diag}(1,-1)(s_1,s_2)^TL^{\ell} \nonumber \\&=\displaystyle{\sum_{\ell=0}^{n-1}}K^{n-\ell-1}r_1s_1^TL^{\ell}-\displaystyle{\sum_{\ell=0}^{n-1}}K^{n-\ell-1}r_2s_2^TL^{\ell} \nonumber \\
&=\partial_{x_n}M_1(r,s)-\partial_{x_n}M_2(r,s),
\label{Derivative of M}
\end{align}
where
\begin{align}\label{def-M1-M2}
	\partial_{x_n}M_1(r,s):=\displaystyle{\sum_{\ell=0}^{n-1}}K^{n-\ell-1}r_1s_1^TL^{\ell},~~~~
	\partial_{x_n}M_2(r,s):=\displaystyle{\sum_{\ell=0}^{n-1}}K^{n-\ell-1}r_2s_2^TL^{\ell}.
\end{align}
The two relations in \eqref{def-M1-M2} can be understood as the following.
Suppose that we have two Sylvester equations,
\begin{align}\label{Syl-eq-expand}
	KM_1-M_1L=r_1s_1^T,~~~~KM_2-M_2L=r_2s_2^T,
\end{align}
while $r=(r_1,r_2)$ and $s=(s_1,s_2)$ satisfy the dispersion relation \eqref{Dispersion relations}. 
Then, following \cite{LQYZ-SAPM-2022,LSS-2023}, we can derive \eqref{def-M1-M2} from \eqref{Syl-eq-expand}.

Therefore, $v$ defined in \eqref{u,v_quasi-Grammian} is a quasi-Grammian solution of the Yang equation [Eq. \eqref{Yang eq_mn}] in the sense that $M_1(r,s)$ and $M_2(r,s)$ are the Grammian-like matrices and so is $M(r,s)$. 

\subsection{An Extension of the Cauchy Matrix Approach ($G=GL(2, \mathbb{C})$)}
In this section, we generalize the  CMA in  Sec. 3.1 by modification of the dispersion relation \eqref{Dispersion relations} in an extended form. 
Let us consider 
the Sylvester equation for $\widetilde{M}(\widetilde{r}, \widetilde{s})\in \mathbb{C}_{N\times N}[x]$,
\begin{eqnarray}
\label{Sylvester eq_tilde}
K\widetilde{M}(\widetilde{r}, \widetilde{s})
-
\widetilde{M}(\widetilde{r}, \widetilde{s})L=\widetilde{r}~\!\widetilde{s}^{T}
\end{eqnarray}
where $\widetilde{r}, \widetilde{s} \in \mathbb{C}_{N\times 2}[x]$ 
satisfy the following differential recurrence relations
\begin{eqnarray}
\label{Differential recurrence_x_j}
\partial_{x_{j+1}}\widetilde{r}=K\partial_{x_{j}}\widetilde{r},~~\partial_{x_{j+1}}\widetilde{s}=L^{T}\partial_{x_{j}}\widetilde{s}, ~j \in \mathbb{Z}.
\end{eqnarray}
 
Solutions of the Sylvester equation [Eq. \eqref{Sylvester eq_tilde}] are also represented by dressed Cauchy matrices. 
Note that any solutions $(\widetilde{r}$, $\widetilde{s})$ of 
the previous dispersion relation \eqref{Dispersion relations} 
satisfy the differential one \eqref{Differential recurrence_x_j}. 
%
Once we set $\partial_{x_0}\widetilde{r}=\widetilde{r}\sigma_3$ and $\partial_{x_0}\widetilde{s}=-\widetilde{s}\sigma_3$, we can get back to \eqref{Dispersion relations} from \eqref{Differential recurrence_x_j}.

Under this setting, our goal is to show that \eqref{u,v_quasi-Grammian} can be generalized to 
\begin{eqnarray}
\label{tilde(u, v)}
\widetilde{u}
=
\left|
\begin{array}{cc}
\widetilde{M}(\widetilde{r}, \widetilde{s}) & \widetilde{r} \\
\widetilde{s}^T & \fbox{$0$}
\end{array}
\right|,~
\widetilde{v}
=
\left|
\begin{array}{cc}
\widetilde{M}(\widetilde{r}, \widetilde{s}) & K^{-1}\widetilde{r} \\
\widetilde{s}^T & \fbox{$I$}
\end{array}
\right|
\end{eqnarray}
so that $\widetilde{v}$ still satisfies the Yang equation [Eq. \eqref{Yang eq_mn}].
Here, we use the tilde notation to distinguish the generalized solution from the original one.

Firstly, we find that the dressed Cauchy matrix $\widetilde{M}(\widetilde{r}, \widetilde{s})$ still has the Grammian-like structure because its derivative can be expressed by the following sum of scalar products:

\newtheorem{Lem_3.1}{Lemma}[section]\label{lemma-3-1}
\begin{Lem_3.1} 
For $K$ and $L$ that do not share eigenvalues, 
under the conditions \eqref{Sylvester eq_tilde} and \eqref{Differential recurrence_x_j}, 
	the Cauchy matrix $\widetilde{M}(\widetilde{r}, \widetilde{s})$ satisfies
	\begin{eqnarray}
	\partial_{x_{j+1}}\widetilde{M}(\widetilde{r}, \widetilde{s})
	=(\partial_{x_j}\widetilde{r})\widetilde{s}^{T} + \left[\partial_{x_j}\widetilde{M}(\widetilde{r}, \widetilde{s})\right]L
	=K\left[\partial_{x_j}\widetilde{M}(\widetilde{r}, \widetilde{s}) \right] - \widetilde{r}(\partial_{x_j}\widetilde{s}^{T}), ~j \in \mathbb{Z}.
	\end{eqnarray}
\end{Lem_3.1}
\textit{Proof:}
The proof is given in Appendix A.
$\hfill\Box$   \medskip  \\ 

Note that $\widetilde{M}(\widetilde{r}, \widetilde{s})$ satisfies the requirement of the following Lemma 3.2 and, therefore, the derivative of \eqref{tilde(u, v)} can be rewritten in the form of \eqref{Derivative formula_grammian}.

\newtheorem{Lem_3.2}[Lem_3.1]{Lemma}
\begin{Lem_3.2}
	{\bf Derivative formula of quasi-Grammian \cite{Gilson-2007}.}
	\medskip \\
	Let $A$ be a $N \times N$ matrix, $B$ be a $N \times 1$ column matrix, $C$ be a $1 \times N$ row matrix, and $d$ be a $1 \times 1$ matrix.
	If the derivative of matrix A can be expressed as
	\begin{eqnarray}
	\partial A=\sum_{\ell=1}^{k}E_{\ell}F_{\ell},
	\end{eqnarray}
	where $E_{\ell}$ and $F_{\ell}$ stand for certain square matrices,
	then we have the following derivative formula of the quasideterminant:
	\begin{eqnarray}
	\label{Derivative formula_grammian}
	&&\partial\left|
	\begin{array}{cc}
	A & B \\
	C & \fbox{$d$}
	\end{array}
	\right|
	= 
	\partial d
	+
	\left|
	\begin{array}{cc}
	A & B \\
	\partial C & \fbox{$0$}
	\end{array}
	\right|
	+
	\left|
	\begin{array}{cc}
	A & \partial B \\
	C & \fbox{$0$}
	\end{array}
	\right|
	+
	\sum_{\ell=1}^{k}
	\left|
	\begin{array}{cc}
	A & E_{\ell} \\
	C & \fbox{$0$}
	\end{array}
	\right|
	\left|
	\begin{array}{cc}
	A & B \\
	F_{\ell} & \fbox{$0$}
	\end{array}
	\right|.
	\end{eqnarray}
\end{Lem_3.2}

Now, we can apply \eqref{Derivative formula_grammian} to verify that 
 $\widetilde{v}$ in   
\eqref{tilde(u, v)} satisfies the Yang equation [Eq. \eqref{Yang eq_mn}]. Our arguments are summarized as the following Theorem 3.3. 

\newtheorem{Thm_3.3}[Lem_3.1]{Theorem}
\begin{Thm_3.3}
Under the conditions \eqref{Sylvester eq_tilde} and \eqref{Differential recurrence_x_j}, 
$\widetilde{u}$ and $\widetilde{v}$ in \eqref{tilde(u, v)} satisfy the differential recurrence relation 
\begin{eqnarray}
(\partial_{x_{j+1}}\widetilde{v})\widetilde{v}^{-1}=\partial_{x_{j}}\widetilde{u}, ~ j \in \mathbb{Z},
\end{eqnarray}
and therefore, such $\widetilde{v}$ is a solution of the Yang equation:
\begin{eqnarray}
\label{Yang eq_x_n}
\partial_{x_{m}}[(\partial_{x_{n+1}}\widetilde{v})\widetilde{v}^{-1}]
-
\partial_{x_{n}}[(\partial_{x_{m+1}}\widetilde{v})\widetilde{v}^{-1}]=0,
\end{eqnarray}
for any given $n, m \in \mathbb{Z}$.
\end{Thm_3.3}
\textit{Proof:} ~
For simplicity, we use $\widetilde{M}$ to denote $\widetilde{M}(\widetilde{r}, \widetilde{s})$ in the following proof.
From Lemma 3.1, Lemma 3.2, and Eq. \eqref{Differential recurrence_x_j}, we have
\begin{eqnarray}
\partial_{x_{j+1}}\widetilde{v}
&\!\!\!\!\!\!=\!\!\!\!\!\!&
\left|
\begin{array}{cc}
\widetilde{M} & K^{-1}\widetilde{r} \\
\partial_{x_{j+1}}\widetilde{s}^{T} & \fbox{$0$}
\end{array}
\right|
+
\left|
\begin{array}{cc}
\widetilde{M} & K^{-1}(\partial_{x_{j+1}}\widetilde{r}) \\
\widetilde{s}^{T} & \fbox{$0$}
\end{array}
\right|
\nonumber \\
&&\!\!\!\!\!\!+
\left|
\begin{array}{cc}
\widetilde{M} & \partial_{x_{j}}\widetilde{r} \\
\widetilde{s}^{T} & \fbox{$0$}
\end{array}
\right|
\left|
\begin{array}{cc}
\widetilde{M} & K^{-1}\widetilde{r} \\
\widetilde{s}^{T} & \fbox{$0$}
\end{array}
\right|
+
\left|
\begin{array}{cc}
\widetilde{M} & \partial_{x_{j}}\widetilde{M} \\
\widetilde{s}^{T} & \fbox{$0$}
\end{array}
\right|
\left|
\begin{array}{cc}
\widetilde{M} & K^{-1}\widetilde{r} \\
L & \fbox{$0$}
\end{array}
\right|
\nonumber \\
&\!\!\!\!\!\!=\!\!\!\!\!\!&
\left|
\begin{array}{cc}
\widetilde{M} & K^{-1}\widetilde{r} \\
\partial_{x_{j}}\widetilde{s}^{T}L & \fbox{$0$}
\end{array}
\!\right|
+
\left|
\begin{array}{cc}
\widetilde{M} & \partial_{x_{j}}\widetilde{r} \\
\widetilde{s}^{T} & \fbox{$0$}
\end{array}
\!\right|
\left|
\begin{array}{cc}
\widetilde{M} & K^{-1}\widetilde{r} \\
\widetilde{s}^{T} & \fbox{$I$}
\end{array}
\!\right|
+
\left|
\begin{array}{cc}
\widetilde{M} & \partial_{x_{j}}\widetilde{M} \\
\widetilde{s}^{T} & \fbox{$0$}
\end{array}
\!\right|
\left|
\begin{array}{cc}
\widetilde{M} & K^{-1}\widetilde{r} \\
L & \fbox{$0$}
\end{array}
\!\right|. ~~~~~~~~
\end{eqnarray}
From the definition of quasideterminants and the Sylvester equation [Eq. \eqref{Sylvester eq}], we have
\begin{eqnarray}
\left|
\begin{array}{cc}
	\widetilde{M} & K^{-1}\widetilde{r} \\
	\partial_{x_{j}}\widetilde{s}^{T}L & \fbox{$0$}
\end{array}
\!\right|
&\!\!\!\!=\!\!\!\!&
\left|
\begin{array}{cc}
\widetilde{M} & (\widetilde{M}L)\widetilde{M}^{-1}K^{-1}\widetilde{r} \\
\partial_{x_{j}}\widetilde{s}^{T} & \fbox{$0$}
\end{array}
\!\right|
=
\left|
\begin{array}{cc}
\widetilde{M} & (K\widetilde{M}-\widetilde{r}~\!\widetilde{s}^{T})\widetilde{M}^{-1}K^{-1}\widetilde{r} \\
\partial_{x_{j}}\widetilde{s}^{T} & \fbox{$0$}
\end{array}
\!\right| ~~~~~~~~
\nonumber \\
&\!\!\!\!=\!\!\!\!&
\left|
\begin{array}{cc}
\widetilde{M} & \widetilde{r} \\
\partial_{x_{j}}\widetilde{s}^{T}  & \fbox{$0$}
\end{array}
\right|
+
\left|
\begin{array}{cc}
\widetilde{M} & \widetilde{r} \\
\partial_{x_{j}}\widetilde{s}^{T} & \fbox{$0$}
\end{array}
\right|
\left|
\begin{array}{cc}
\widetilde{M} & K^{-1}\widetilde{r} \\
\widetilde{s}^{T} & \fbox{$0$}
\end{array}
\right|
\nonumber \\
&\!\!\!\!=\!\!\!\!&
\left|
\begin{array}{cc}
\widetilde{M} & \widetilde{r} \\
\partial_{x_{j}}\widetilde{s}^{T}  & \fbox{$0$}
\end{array}
\right|
\left|
\begin{array}{cc}
\widetilde{M} & K^{-1}\widetilde{r} \\
\widetilde{s}^{T} & \fbox{$I$}
\end{array}
\right|. 
\end{eqnarray}
Through similar calculation, we have
\begin{eqnarray}
\left|
\begin{array}{cc}
\widetilde{M} & K^{-1}\widetilde{r} \\
L & \fbox{$0$}
\end{array}
\right|
=
\left|
\begin{array}{cc}
\widetilde{M} & \widetilde{r} \\
I  & \fbox{$0$}
\end{array}
\right|
\left|
\begin{array}{cc}
\widetilde{M} & K^{-1}\widetilde{r} \\
\widetilde{s}^{T} & \fbox{$I$}
\end{array}
\right|.
\end{eqnarray}
Now we can conclude that
\begin{eqnarray}
\partial_{x_{j+1}}\widetilde{v}
&\!\!\!\!=\!\!\!\!&
\left\{
\left|
\begin{array}{cc}
\widetilde{M} & \widetilde{r} \\
\partial_{x_{j}}\widetilde{s}^{T} & \fbox{$0$}
\end{array}
\right|
+
\left|
\begin{array}{cc}
\widetilde{M} & \partial_{x_{j}}\widetilde{r} \\
\widetilde{s}^{T} & \fbox{$0$}
\end{array}
\right|
+
\left|
\begin{array}{cc}
\widetilde{M} & \partial_{x_{j}}\widetilde{M} \\
\widetilde{s}^{T} & \fbox{$0$}
\end{array}
\right|
\left|
\begin{array}{cc}
\widetilde{M} & \widetilde{r} \\
I & \fbox{$0$}
\end{array}
\right|
\right\}
\left|
\begin{array}{cc}
\widetilde{M} & K^{-1}\widetilde{r} \\
\widetilde{s}^{T} & \fbox{$I$}
\end{array}
\right| ~~~~~~~~
\nonumber \\
&\!\!\!\!=\!\!\!\!&
\left\{
\partial_{x_{j}}
\left|
\begin{array}{cc}
\widetilde{M} & \widetilde{r} \\
\widetilde{s}^{T} & \fbox{$0$}
\end{array}
\right|
\right\}
\left|
\begin{array}{cc}
\widetilde{M} & K^{-1}\widetilde{r} \\
\widetilde{s}^{T} & \fbox{$I$}
\end{array}
\right|
=(\partial_{x_{j}}\widetilde{u})\widetilde{v}. 
\end{eqnarray}
Here, we have used Lemma 3.2 again with respect to $\widetilde{u}$.
$\hfill\Box$ 

\subsection{Unitary Solutions in the WZW$_4$ Model}
For the physical purpose of clarifying the WZW$_4$ model, in this section we aim to find the unitary solutions of \eqref{Yang eq_x_n} on the Ultrahyperbolic space $\mathbb{U}$ whose metric can be realized by \eqref{Reality condition_U}. Let us assign the coordinates $(x_{n+1}, x_{m}, x_{m+1}, x_{n})$ in \eqref{Yang eq_x_n} to $(z, \widetilde{z}, w, \widetilde{w}) \in \mathbb{R}^4$. 
The differential recurrences in \eqref{Differential recurrence_x_j} now reduce to
\begin{eqnarray}
\label{Differential recurrence_zw}
\left\{
\begin{array}{l}
\partial_{z}\widetilde{r}=K\partial_{\widetilde{w}}\widetilde{r}
\\
\partial_{w}\widetilde{r}=K\partial_{\widetilde{z}}\widetilde{r}
\end{array},
\right. ~~
\left\{
\begin{array}{l}
\partial_{z}\widetilde{s}=L^{T}\partial_{\widetilde{w}}\widetilde{s}
\\
\partial_{w}\widetilde{s}=L^{T}\partial_{\widetilde{z}}\widetilde{s}
\end{array}, ~z, \widetilde{z}, w, \widetilde{w} \in \mathbb{R}, 
\right.
\end{eqnarray}
which can, in fact, be regarded as a special class of the linear systems in the context of the binary Darboux transformations \cite{Nimmo-2000},
\begin{eqnarray}
\label{Linear systems of ASDYM}
\begin{array}{l}
\left\{
\begin{array}{l}
\left[\partial_{z} - (\partial_{z}J)J^{-1} \right]\theta -(\partial_{\widetilde{w}}\theta)\Lambda = 0,
\smallskip \\
\left[\partial_{w} - (\partial_{w}J)J^{-1} \right]\theta -(\partial_{\widetilde{z}}\theta)\Lambda = 0,
\end{array}
\right.
\medskip \\
\left\{
\begin{array}{l}
\left[\partial_{z} - (\partial_{z}J^{-\dagger})J^{\dagger} \right]\rho -(\partial_{\widetilde{w}}\rho)\Xi = 0,
\smallskip \\
\left[\partial_{w} - (\partial_{w}J^{-\dagger})J^{\dagger} \right]\rho -(\partial_{\widetilde{z}}\rho)\Xi = 0,
\end{array}
\right.
\end{array}
~~z, \widetilde{z}, w, \widetilde{w} \in \mathbb{R}, 
~J^{-\dagger}:=(J^{-1})^{\dagger},
\end{eqnarray}
which compose the Lax pair of the  Yang   equation. 
This coincides with \eqref{Differential recurrence_zw} 
through the following identification:
\begin{eqnarray}
J= I,~ (\theta, \Lambda)=(\widetilde{s}^{T}, L),~(\rho, \Xi)=(\widetilde{r}^{\dagger},  K^{\dagger}).
\end{eqnarray}
The linear systems in \eqref{Linear systems of ASDYM}
in the case of $J=I$ are used as initial linear systems when the binary Darboux transformations are applied. Therefore, we have the following theorem on the Ultrahyperbolic space $\mathbb{U}$:
\newtheorem{Thm_3.4}[Lem_3.1]{Theorem}
\begin{Thm_3.4} {\bf (cf. \cite{Nimmo-2000})}
	Let us consider the Sylvester equation:
	\begin{eqnarray}
	 \label{Sylvester eq_Omega}
	\Xi^{\dagger}\Omega(
	\theta, \rho) - \Omega(\theta, \rho)\Lambda = \rho^{\dagger}\theta,
	\end{eqnarray}
	where $(\theta, \Lambda)$ and  $(\rho, \Xi)$ satisfy the linear systems in \eqref{Linear systems of ASDYM}.
	If we define $\hat{U}$ and $\hat{J}$ by
	\begin{eqnarray}
	\hat{U}
	:= \left|
	\begin{array}{cc}
	\Omega(\theta, \rho) & \rho^{\dagger} 
	\smallskip \\
	\theta  & \fbox{$U$}
	\end{array}
	\right|, ~~~
	\hat{J}
	:= \left|
	\begin{array}{cc}
	\Omega(\theta, \rho) & \Xi^{-\dagger}\rho^{\dagger} 
	\smallskip \\
	\theta  & \fbox{$I$}
	\end{array}
	\right|J,
        \label{J_nonunitary}
	\end{eqnarray}
	where $U$ satisfies 
	\begin{eqnarray}
	\partial_{\widetilde{z}}U=(\partial_{w}J)J^{-1}, ~~\partial_{\widetilde{w}}U=(\partial_{z}J)J^{-1},
	\end{eqnarray}
	then $\hat{U}$ and $\hat{J}$ satisfy the differential relations
	\begin{eqnarray}
	\partial_{\widetilde{z}}\hat{U}=(\partial_{w}\hat{J})\hat{J}^{-1}, ~~\partial_{\widetilde{w}}\hat{U}=(\partial_{z}\hat{J})\hat{J}^{-1}.
	\end{eqnarray}
	That is, $\hat{J}$ satisfies the Yang equation
	\begin{eqnarray}
	\partial_{\widetilde{z}}\left[ (\partial_{z}\hat{J})\hat{J}^{-1} \right]
	-
	\partial_{\widetilde{w}}\left[ (\partial_{w}\hat{J})\hat{J}^{-1} \right]=0
	\end{eqnarray}
	on the Ultrahyperbolic space $\mathbb{U}$.
\end{Thm_3.4}
\textit{Proof.} 
We can check that Lemma 3.1 still holds for linear systems in \eqref{Linear systems of ASDYM}, that is, the Cauchy matrix $\Omega$ satisfies the following relations:
\begin{eqnarray}
\label{Omega_derivative}
\left\{
\begin{array}{l}
\partial_{z}\Omega
=\Xi^{\dagger}(\partial_{\widetilde{w}}\Omega)-\rho^{\dagger}(\partial_{\widetilde{w}}\theta)
=(\partial_{\widetilde{w}}\Omega)\Lambda + (\partial_{\widetilde{w}}\rho^{\dagger})\theta
\smallskip \\
\partial_{w}\Omega
=\Xi^{\dagger}(\partial_{\widetilde{z}}\Omega)-\rho^{\dagger}(\partial_{\widetilde{z}}\theta)
=(\partial_{\widetilde{z}}\Omega)\Lambda + (\partial_{\widetilde{z}}\rho^{\dagger})\theta
\end{array}.
\right.
\end{eqnarray}
Since the proof is quite similar to the proof in Theorem 3.3, we just skip all the details here.
$\hfill\Box$\\

Note that if we choose $J$ to be unitary and $(\rho, \Xi)=(\theta, \Lambda)$, the second system of equations of \eqref{Linear systems of ASDYM} is now identical to the first one, that is, $(\theta, \Lambda)$ satisfies the following linear system:
\begin{eqnarray}
\label{Linear systems of ASDYM_unitary}
\left\{
\begin{array}{l}
\left[\partial_{z} - (\partial_{z}J)J^{-1} \right]\theta -(\partial_{\widetilde{w}}\theta)\Lambda = 0
\smallskip \\
\left[\partial_{w} - (\partial_{w}J)J^{-1} \right]\theta -(\partial_{\widetilde{z}}\theta)\Lambda = 0
\end{array}.
\right.
\end{eqnarray}
Under this setting, we can obtain a class of unitary solutions as the following corollary: 
\newtheorem{Cor_3.5}[Lem_3.1]{Corollary}
\begin{Cor_3.5}
Let $J$ be unitary and $\Omega(\theta, \theta)$ be the  dressed Cauchy matrix satisfying the Sylvester equation
\begin{eqnarray}
\label{Sylvester eq_hermitian}
\Lambda^{\dagger}\Omega(\theta, \theta) - \Omega(\theta, \theta)\Lambda=\theta^{\dagger}\theta.
\end{eqnarray}
Then, we can obtain unitary solutions in the form of 
\begin{eqnarray}
\label{J_unitary}
\hat{J}
=
\left|
\begin{array}{cc}
\Omega(\theta, \theta) & \Lambda^{-\dagger}\theta^{\dagger}
\\
\theta & \fbox{$I$}
\end{array}
\right|J.
\end{eqnarray}
\end{Cor_3.5}
\textit{Proof:} 
By using the fact that $\Omega^{\dagger}=-\Omega$ [see Eq. \eqref{Sylvester eq_hermitian}] and direct calculation, we have
\begin{eqnarray}
\hat{J}^{\dagger}\hat{J}
&\!\!\!\!=\!\!\!\!& J^{\dagger}(I-\theta\Lambda^{-1}\Omega^{-\dagger}\theta^{\dagger})(I-\theta\Omega^{-1}\Lambda^{-\dagger}\theta^{\dagger})J  \nonumber \\
&\!\!\!\!=\!\!\!\!&
J^{\dagger}[I - \theta\Lambda^{-1}\Omega^{-\dagger}\theta^{\dagger}-\theta\Omega^{-1}\Lambda^{-\dagger}\theta^{\dagger} + \theta\Lambda^{-1}\Omega^{-\dagger}
(\Lambda^{\dagger}\Omega - \Omega\Lambda)\Omega^{-1}\Lambda^{-\dagger}\theta^{\dagger}]J
\nonumber \\
&\!\!\!\!=\!\!\!\!&
J^{\dagger}
[I 
-\theta(I + \Lambda^{-1}\Omega^{-\dagger}\Omega\Lambda)\Omega^{-1}\Lambda^{-\dagger}\theta^{\dagger}]J
=J^{\dagger}J
=I.
\end{eqnarray}
Similarly, we also have $JJ^{\dagger}=I$. Therefore, $\hat{J}$ 
is unitary.
$\hfill\Box$ 

A strategy to construct exact solutions is summarized as follows. Firstly, we solve the initial linear systems to get $(\theta, \Lambda)$,
\begin{eqnarray}
\label{initial}
\partial_{z}\theta = (\partial_{\widetilde{w}}\theta)\Lambda,~~~
\partial_{w} \theta = (\partial_{\widetilde{z}}\theta)\Lambda.
\end{eqnarray}
The solutions $(\theta, \Lambda)$ are called the input data. 
Secondly, we solve the Sylvester equation [Eq. \eqref{Sylvester eq_hermitian}] 
by the input data
$(\theta,\Lambda)$ to get $\Omega$. Finally, we  can get a unitary solution $\hat{J}$ via \eqref{J_unitary}. 


\section{Exact Solitons in the WZW$_4$ Model}


In this section, we give several exact soliton solutions in the WZW$_4$ model and calculate the corresponding action densities of the NL$\sigma$M term for  one-solitons in  Sec. \ref{sec-4-0} and for two-solitons in Sec. \ref{sec-4-1}.
We clarify that the quasi-Wronskian solution \cite{Hamanaka-2023} and quasi-Grammian solutions \eqref{J_unitary} give the same action densities for one- and two-soliton solutions. In  Sec. \ref{sec-4-2}, we discuss some limits of the two-soliton solutions, some of which lead to resonance solutions. 

\subsection{Exact One-Solitons}\label{sec-4-0}

A set of input data $(\theta,\Lambda)$ for one-solitons can be given by solving \eqref{initial},
\begin{equation}
	\label{1ss}
	\theta :=
	\left(
	\begin{array}{c}
		a_{1}^2e^{\xi_{1}} 
		\\
		b_{1}^2e^{-\xi_{1}} 
	\end{array}
	\right),~~
	\Lambda:=\left(\lambda_1\right),
\end{equation}
where
\begin{equation}
	\xi_{1}:=\lambda_{1}\alpha_{1}z + \beta_{1}\widetilde{z} + \lambda_{1}\beta_{1}w + \alpha_{1}\widetilde{w}, ~ a_{1}, b_{1}, \lambda_{1}, \alpha_{1}, \beta_{1} \in \mathbb{C}.
\end{equation}
The corresponding solution of the Sylvester equation is 
\begin{eqnarray}
\label{Omega_jk_1ss}
\Omega
=\frac{\left|a_1\right|^2 e^{\overline{\xi}_{1}+ \xi_{1}} + \left|b_1 \right|^2 e^{-(\overline{\xi}_{1}+ \xi_{1})}}{\overline{\lambda}_{1} - \lambda_{1}},
\end{eqnarray}
where $\overline\lambda$ stands for the complex conjugate of $\lambda$.
The resulting NL$\sigma$M action density can be calculated via \eqref{J_unitary} as
\begin{equation}\label{NLsigmaM-1SS}
	{\cal{L}}_\sigma=-\frac{1}{16\pi}
	\mbox{Tr}\left[(\partial_{\mu}\hat{J})\hat{J}^{-1}(\partial^{\mu}\hat{J})\hat{J}^{-1}\right] 
	=\frac{d_{11}}{8\pi}\mathrm{sech}^2\widetilde{X}_1,
\end{equation}
where
\begin{subequations}
	\begin{align}
		&d_{11}=\frac{(\alpha_1\overline\beta_1-\beta_1\overline\alpha_1)(\lambda_1-\overline\lambda_1)^3}{|\lambda_1|^2}, \\
		&\widetilde{X}_1=X_1+\log\delta_1:=\xi_1+\overline\xi_1+\log\delta_1,~~\delta_1=|a_1|^2/|b_1|^2.
	\end{align}
\end{subequations}
The result of ${\cal{L}}_\sigma$ is exactly the same as the quasi-Wronskian one-soliton solution in \cite{Hamanaka-2023}, where the input data $\theta$ and $\Lambda$ are both $2\times 2$ matrices rather than \eqref{1ss}.

The resulting Wess-Zumino action density ${\cal{L}}_{\scriptsize{\mbox{WZ}}}$ identically vanishes \cite{Hamanaka-2023}. Hence, the WZW$_4$ action density is the same as ${\cal{L}}_\sigma$ in \eqref{NLsigmaM-1SS}.

\subsection{Exact Two-Solitons}\label{sec-4-1}

A set of input data for two-solitons can be given by the following 2 $\times$ 2 matrix pair $(\theta, \Lambda)$: 
\begin{eqnarray}
\label{2ss-matrix-pair}
\theta
:=
\left(
\begin{array}{cc}
a_{1}^2e^{\xi_{1}} & a_{2}^2e^{\xi_{2}}
\\
b_{1}^2e^{-\xi_{1}} & b_{2}^2e^{-\xi_{2}}
\end{array}
\right),~~
\Lambda
:=
\left(
\begin{array}{cc}
\lambda_{1} & 0
\\
0 & \lambda_{2}
\end{array}
\right), 
\end{eqnarray}
where
\begin{eqnarray}
\label{Xi}
\xi_{j}:=\lambda_{j}\alpha_{j}z + \beta_{j}\widetilde{z} + \lambda_{j}\beta_{j}w + \alpha_{j}\widetilde{w}, ~~~ a_{j}, b_{j}, \lambda_{j}, \alpha_{j}, \beta_{j} \in \mathbb{C}, ~ j=1, 2.
\end{eqnarray}
The NL$\sigma$M action density can be calculated as (see Appendix \ref{Calculation of Two-Soliton} for details)
\begin{eqnarray}
{\cal{L}}_\sigma&=&
-\frac{1}{16\pi}
\mbox{Tr}\left[(\partial_{\mu}\hat{J})\hat{J}^{-1}(\partial^{\mu}\hat{J})\hat{J}^{-1}\right] \nonumber  \\
&\!\!\!\! = \!\!\!\!&
\label{NL Sigma term_2-Soliton_form 2}
\frac{
	\left\{ 
	\begin{array}{l}
	~c_{1}c_{2}
	\displaystyle{\left[
		d_{11}\cosh^2\widetilde{X}_{2} + d_{22}\cosh^2\widetilde{X}_{1} 
		\right]
	}
	\medskip \\
	\!\!+ c_{1}c_{3}
	\left[
	\displaystyle{
		d_{12} \cosh^2\left( 
		\frac{\widetilde{X}_{1} + \widetilde{X}_{2} - i\widetilde{\Theta}_{12}}{2}
		\right)
		
		+ \overline{d}_{12}\cosh^2\left( 
		\frac{\widetilde{X}_{1} + \widetilde{X}_{2} + i\widetilde{\Theta}_{12}}{2}
		\right)
	}
	\right]
	\medskip \\
	\!\!- c_{2}c_{3}\left[
	\displaystyle{
		f_{12}~\!\sinh^2\left(
		\frac{\widetilde{X}_{1} - \widetilde{X}_{2} -i\widetilde{\Theta}_{12}}{2}
		\right)
		+ \overline{f}_{12}~\!\sinh^2\left(
		\frac{\widetilde{X}_{1} - \widetilde{X}_{2} +i\widetilde{\Theta}_{12}}{2}
		\right)
	}
	\right]
	\end{array}
	\!\!\!\right\}
}
{2\pi\left[
	c_{1}\cosh(
	\widetilde{X}_{1} + \widetilde{X}_{2} 
	)
	+ c_{2}\cosh(
	\widetilde{X}_{1} - \widetilde{X}_{2}
	)
	+ c_{3}\cos\widetilde{\Theta}_{12}
	\right]^2},~~~~~~
\end{eqnarray}
where
\begin{subequations}
\begin{eqnarray}
c_{1}&\!\!\!\!:=\!\!\!\!&(\lambda_{1}-\lambda_{2})(\overline{\lambda}_{1}-\overline{\lambda}_{2}),~
c_{2}:=(\lambda_{1}-\overline{\lambda}_{2})(\overline{\lambda}_{1}-\lambda_{2}),~
c_{3}:=(\lambda_{1}-\overline{\lambda}_{1})(\lambda_{2}-\overline{\lambda}_{2}),
\\
d_{jk}&\!\!\!\!:=\!\!\!\!&  \label{d_jk}
\frac{(\alpha_{j}\overline{\beta}_{k} - \beta_{j}\overline{\alpha}_{k})(\lambda_{j} - \overline{\lambda}_{k})^3}{\lambda_{j}\overline{\lambda}_{k}},~
f_{jk}:=
\frac{(\alpha_{j}\beta_{k} - \beta_{j}\alpha_{k})(\lambda_{j} - \lambda_{k})^3}
{\lambda_{j}\lambda_{k}}, ~j, k =1, 2,
\\
\label{Xj}
\widetilde{X}_{j}&\!\!\!\!:=\!\!\!\!&  X_{j} + \log \delta_{j}:=\xi_{j} + \overline{\xi}_{j} + \log \delta_{j}, ~\delta_{j}:=|a_{j}|^2 / |b_j|^2,
\\
\widetilde{\Theta}_{12}&\!\!\!\!:=\!\!\!\!&\Theta_{1} - \Theta_{2} + \phi, ~
\Theta_{j}:=-i(\xi_{j}-\overline{\xi}_{j}), ~j=1, 2,
\\
\phi&\!\!\!\!:=\!\!\!\!&2\mbox{Arg}(a_{1}\overline{a}_{2} / b_{1}\overline{b}_{2})
=2\mbox{Arg}(a_{1}b_{2} / a_{2}b_{2}).
\end{eqnarray}
\end{subequations}
The resulting ${\cal{L}}_\sigma$ is exactly the same as the case of the quasi-Wronskian two-soliton solution in \cite{Hamanaka-2023}. 
The input data here are much simpler than those in \cite{Hamanaka-2023} because the input data in \cite{Hamanaka-2023} consist of two sets of  $2\times 2$ matrix pairs, $(\theta_j, \Lambda_j), j=1,2$.

Note that the NL$\sigma$M action density ${\cal{L}}_\sigma$ is non-singular and real-valued. The distribution of ${\cal{L}}_\sigma$ behaves like KP two-solitons
because we can show that the asymptotic limits of $ {\cal{L}}_\sigma$ are
\begin{eqnarray}
	\label{Asymptotic_2-Soliton}
	-8\pi {\cal{L}}_\sigma 
	\longrightarrow 
	\left\{
	\begin{array}{l}
		d_{11}~\!{\mathrm{sech}^2( \widetilde{X}_1 \pm \widetilde{\delta} )}  ~~~{\mbox{if}}~X_1 ~\mbox{is finite and} ~X_2  \rightarrow  \pm\infty
		\medskip \\
		d_{22}~\!{\mathrm{sech}^2( \widetilde{X}_2 \pm \widetilde{\delta} )}~~~{\mbox{if}}~X_2 ~\mbox{is finite and} ~X_1  \rightarrow  \pm\infty
		
	\end{array},
	\right.
\end{eqnarray}
where the  phase-shift factor is given by  
\begin{eqnarray}\label{Phase_Shift_Factor}
	\widetilde{\delta}:=\displaystyle{\frac{1}{2}~\!\log\left(\frac{c_1}{c_2}\right)} 
	= \log \frac{\left|\lambda_1-\lambda_2\right|}{\left|\lambda_1-\overline{\lambda}_2\right|}.
\end{eqnarray}

\begin{figure}[ht]
	\centering
	\subfigure[]{
		\includegraphics[width=0.3\textwidth]{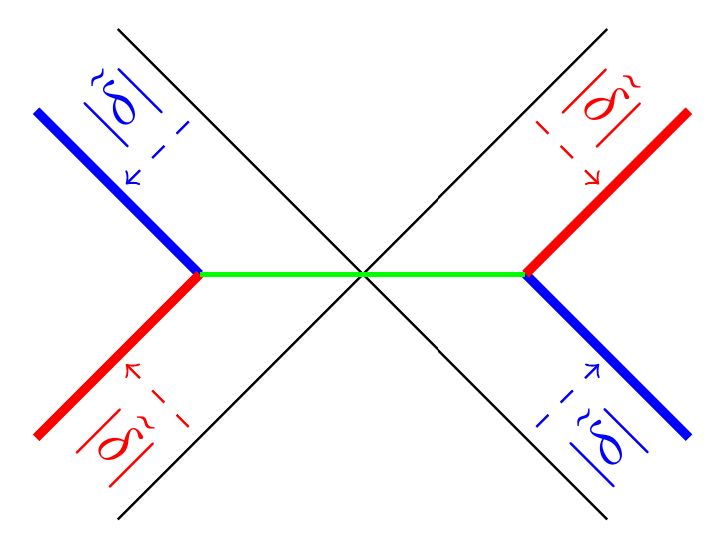}}
	\hspace{1.5cm} 
	\subfigure[]{
		\includegraphics[width=0.22\textwidth]{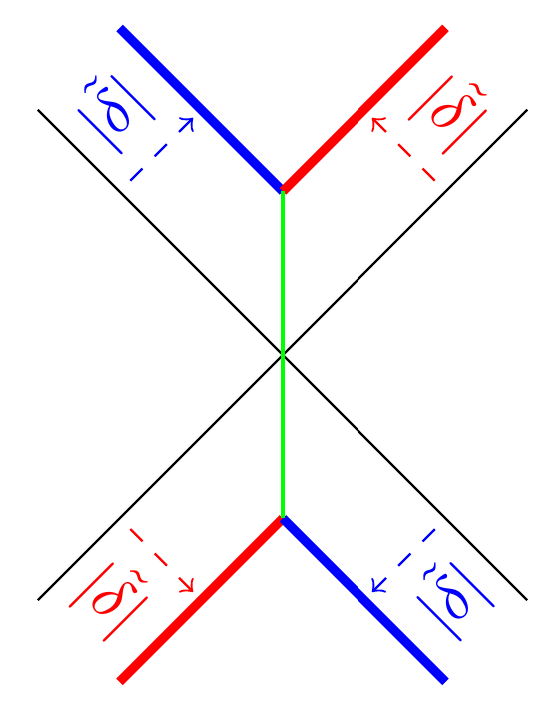}}
	\caption{2D slice  of two-soliton NL$\sigma$M action density 
for $(w, \widetilde{w})=(0, 0)$. 
(a) The case of $\tilde\delta>0$. (b) The case of $\tilde\delta<0$.}
	\label{fig-0}
\end{figure}

\begin{figure}[ht]
	\centering
	\subfigure[]{
		\includegraphics[width=0.4\textwidth]{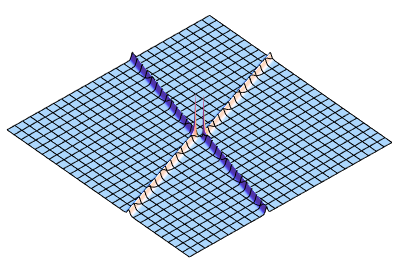}}
	\hspace{0.5cm} 
	\subfigure[]{
		\includegraphics[width=0.4\textwidth]{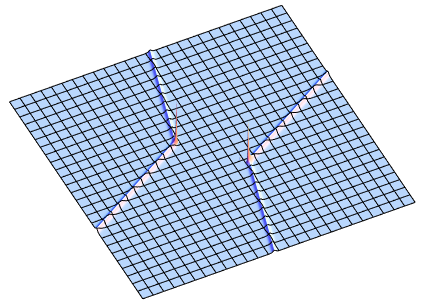}}
	\caption{Plots of the 2D slice of V-shape NL$\sigma$M action density with $\alpha_1=\alpha_2=0.5-0.7i,\beta_1=-0.5-1.5i,\beta_2=1.5+0.5i$, $(z,\tilde z)\in[-20,20]\times[-20,20]$ and $(w,\tilde w)=(0,0)$. (a) The case of $\lambda_1=1+2i$ and $\lambda_2=1-1.5i$. (b) The case of $\lambda_1=1+2i$ and $\lambda_2=1-1.9999999i$.}
	\label{fig-4}
\end{figure}

In  Fig. \ref{fig-0} [also refer to Fig. \ref{fig-4}(b)], we can see the physical meaning of phase shift.  
All figures in this paper are 2D slice plotted by taking $(w, \widetilde{w}) =(0, 0)$.   

The red line and blue line denote  two-line solitons and, as we can see, the V-shape soliton is composed of a half red line soliton and half blue line soliton. The green line describes the distance between the vertices of the two V-shape solitons, which becomes longer and longer as $\tilde\delta$ increases.

We note that when $\lambda_1$ is real, that is, $\lambda_1=\overline{\lambda}_1$, the two-soliton reduces to one-soliton case: 
$\mathcal{L}_\sigma=\displaystyle{\frac{d_{11}}{8\pi}}\mathrm{sech}^2\widetilde{X}_1$ because of 
$c_{3}=d_{22} =0$. 
Similarly, when $\lambda_2$ is real,
$\mathcal{L}_\sigma=\displaystyle{\frac{d_{22}}{8\pi}}\mathrm{sech}^2\widetilde{X}_2$ because of $c_{3}=d_{11} =0$.

\subsection{Resonance Limits of Two-Solitons}\label{sec-4-2}

Let us discuss some limits of the two-solitons where the absolute value of phase shift \eqref{Phase_Shift_Factor} goes to infinity, that is, the cases of (1) $\lambda_{2} \rightarrow \lambda_{1}$ and (2) $\lambda_{2} \rightarrow \overline{\lambda}_{1}$. 
These processes usually lead to resonance states.  

Naive discussion of cases (1) and (2) shows that
\begin{eqnarray}
\label{Reduced solutions of 2-soliton}	
\cal{L}_{\sigma}
\rightarrow
\left\{
\begin{array}{l}
(1):~ 0, ~~\mbox{if} ~~\lambda_{2} \rightarrow \lambda_{1} ~(c_{1}, f_{12} \rightarrow 0)
\medskip \\
(2):~ 0, ~~\mbox{if} ~~\lambda_{2} \rightarrow \overline{\lambda}_{1} ~(c_{2}, d_{12} \rightarrow 0)
\end{array}
\right.
\end{eqnarray}
which is correct for almost all cases because the numerator of \eqref{NL Sigma term_2-Soliton_form 2} tends to zero. 
However, 
we need to be more careful when dealing with the case when the denominator becomes zero. 
For the case (1), there is a particular choice for the parameters: 
$\alpha_{1}=\alpha_{2}:=\alpha, \beta_{1}=\beta_{2}:=\beta$. 
In this case, the  two-line solitons have equal ``amplitude'': $d_{11}=d_{22}$ [cf. Eq. \eqref{Asymptotic_2-Soliton}]. 
For simplicity, we assume the initial phase terms $\text{log}\delta_j$ and $\phi$ to be zero [$i.e., ~a_{j}=b_{j}, j=1, 2$ in \eqref{NL Sigma term_2-Soliton_form 2}]. 
By using the fact that $c_{1}=c_{2}+c_{3}$ and some properties of hyperbolic functions,
the denominator of \eqref{NL Sigma term_2-Soliton_form 2} can be rewritten as 
\begin{eqnarray}
\label{Denominator_2ss}
2\pi\left[
c_{2}\mbox{cosh}(\xi_{1} + \overline{\xi}_{1})\mbox{cosh}(\xi_{2} + \overline{\xi}_{2})
+ c_{3}\mbox{cosh}(\xi_{1} + \overline{\xi}_{2})\mbox{cosh}(\overline{\xi}_{1}+\xi_{2})
\right]^2. 
\end{eqnarray} 
Clearly, \eqref{Denominator_2ss} tends to zero as $\lambda_{2} \rightarrow \lambda_{1} ~ (c_{1} \rightarrow 0 \Rightarrow c_{2} \rightarrow   -c_{3}, \mathbb{\xi}_{1} \rightarrow \mathbb{\xi}_{2})$. Now, we find that $\cal{L}_{\sigma} \rightarrow 0/0 $ in the above case (1). 
Dividing a common factor $c_{1}^2$ in both the denominator and numerator of \eqref{NL Sigma term_2-Soliton_form 2} and using some properties of hyperbolic functions, the denominator \eqref{Denominator_2ss}/$c_{1}^2$ can be rewritten as
\begin{eqnarray}
\label{Denominator_2ss_form 2}
2\pi\left[
\begin{array}{l}
\mbox{cosh}^2X_{1}|\cosh\xi|^2 -\mbox{cosh}X_{1}
\mbox{sinh}X_{1}\mbox{sinh}(\xi + \overline{\xi})
\smallskip \\
+(c_{2}\mbox{cosh}^2X_{1} + c_{3}\mbox{sinh}^2X_{1})
\displaystyle{\left|\frac{\mbox{sinh}\xi}{\lambda}\right|^2}
\end{array}
\right]^2,
\end{eqnarray}
where for fixed $\lambda_{1}$ and $\xi_{1}$, 
we define $\lambda:= \lambda_{2}- \lambda_{1}, \xi:=\xi_{2}-\xi_{1}=\lambda(\alpha z + \beta w).$
Note that as $\lambda \rightarrow 0$, we have
\begin{subequations}	
\begin{eqnarray}
&&\mbox{cosh}\xi \rightarrow 1, ~
\mbox{sinh}(\xi + \overline{\xi}) \rightarrow 0, ~
\\
&&c_{2}\mbox{cosh}^2X_{1} + c_{3}\mbox{sinh}^2X_{1}
\rightarrow
|\lambda_1 - \overline{\lambda_{1}}|^2(\mbox{cosh}^2X_{1} - \mbox{sinh}^2X_{1}) = |\lambda_1 - \overline{\lambda_{1}}|^2,
\\
&& \frac{\mbox{sinh}\xi}{\lambda}
\rightarrow
\dot{\xi}\mbox{cosh}\xi\Big|_{\lambda=0}=\dot{\xi}, ~~~\dot{\xi}:=\partial_{\lambda}\xi = \alpha z + \beta w.
\end{eqnarray}
\end{subequations}
Therefore, \eqref{Denominator_2ss_form 2} converges to 
\begin{eqnarray}
\label{Denominator_limit}
2\pi
\left[
|\lambda_{1} - \overline{\lambda}_{1}|^2|\dot{\xi}_{1}|^2 + \mbox{cosh}^2X_{1}
\right]^2
 = \frac{\pi}{2}\left[
1+2|\lambda_{1} - \overline{\lambda}_{1}|^2|\dot{\xi}_{1}|^2 + \mbox{cosh}(2X_{1})\right]^2
\end{eqnarray}
as $\lambda_{2} \rightarrow \lambda_{1}$, where $ \dot{\xi}_{1}:=\partial_{\lambda_{1}}\xi_{1} = \alpha z + \beta w$. 
As we will see later in the next section, \eqref{Denominator_limit} matches with the denominator of \eqref{Double-pole_action density} exactly. In  Sec. 5, we will focus on solutions of this type ($\lambda_{2} \rightarrow \lambda_{1}$) and construct them systematically.  

Finally, let us discuss the intermediate state described by green lines in Fig. \ref{fig-0} in the setting where $\hat{J}$ is not unitary, which implies that  ${\cal{L}}_\sigma$ are not real-valued in general, even in the resonance limits.

In this case, we have to consider the unreduced non-unitary 
solution \eqref{J_nonunitary} in Theorem 3.4
[not the unitary solution \eqref{J_unitary} in Corollary 3.5],
where there are additional spectral parameters $\mu_1$ and $\mu_2$ 
which come from $\Xi$ (see Appendix \ref{Non-unitary 2-soliton} for details). Then, the vanishing asymptotics in the cases (1) and (2) of \eqref{Reduced solutions of 2-soliton} become nontrivial due to the additional degree of freedom. 
Under the assumption $\Theta_{12}=2n\pi$ or $(2n+1)\pi$, 
the infinite  phase-shift limits are 
\begin{eqnarray}
	\label{Limit solutions_nonunitary}
	\label{Reduced solutions of 2-soliton_nonunitary}	
	\cal{L}_{\sigma}
	\rightarrow
	\left\{
	\begin{array}{l}
		(1)~{\mbox{When}}~\Theta_{12}=(2n+1)\pi:
		\smallskip \\
		~~~~\displaystyle{\frac{-\widetilde{f}_{12}}{8\pi}\mbox{sech}^2(\frac{\widetilde{X}_{1} - \widetilde{X}_{2}}{2})
	}, ~~\mbox{if} ~~\lambda_{2} \rightarrow \lambda_{1} ~(c_{1}, f_{12} \rightarrow 0),
\smallskip \\
		~~~~\displaystyle{\frac{-f_{12}}{8\pi}\mbox{sech}^2(\frac{\widetilde{X}_{1} - \widetilde{X}_{2}}{2})
		}, ~~\mbox{if} ~~\mu_{2} \rightarrow \mu_{1} ~(c_{1}, \widetilde{f}_{12} \rightarrow 0),
		\medskip \\
		(2)~{\mbox{When}}~\Theta_{12}=2n\pi:
		\smallskip \\
		~~~~\displaystyle{
		\frac{d_{12}}{8\pi}\mbox{sech}^2(\frac{\widetilde{X}_{1} +\widetilde{X}_{2}}{2})}, ~~\mbox{if} ~~\lambda_{2} \rightarrow \mu_{1} ~(c_{2}, \widetilde{d}_{12} \rightarrow 0),
 \smallskip \\
 ~~~~\displaystyle{
 	\frac{\widetilde{d}_{12}}{8\pi}\mbox{sech}^2(\frac{\widetilde{X}_{1} +\widetilde{X}_{2}}{2})}, ~~\mbox{if} ~~\mu_{2} \rightarrow \lambda_{1} ~(c_{2}, d_{12} \rightarrow 0).
	\end{array}
	\right.
\end{eqnarray}
Hence, in the infinite phase-shift limit, we can find the intermediate soliton states (described by green lines in Fig.\ref{fig-0}) in the expected directions. Therefore, in the  complex valued settings, solutions of this type suggest the existence of Y-shape solitons \cite{Ohkuma-Wadati-1983}, which are building blocks in the classification of KP line solitons \cite{Kodama-2017, Kodama-Williams-2014}. 

On the other hand, in the real-valued settings, this intermediate soliton  state  vanishes for the NL$\sigma$M action density, as mentioned in \eqref{Reduced solutions of 2-soliton}, and for the Wess-Zumino action density as well, as is seen in Appendix D.4 in \cite{Hamanaka-2023} [where (D.18) and (D.19) vanish in the $\lambda_1\rightarrow \overline{\lambda}_2$ limit; then, $b=0$ and $\mathcal{D}_{12}=\mathcal{D}_{21}=\mathcal{d}_{12}=\mathcal{d}_{21}= 0$, see also (D.15)]. This is a reasonable result because this intermediate state describes a one-soliton distribution, and the Wess-Zumino action density identically vanishes, as mentioned in  Sec. 4.1. Hence, the effect of the Wess-Zumino action density appears only around the two vertices of two V-shape solitons.

\subsection{Multi-Soliton Solutions}

In the same way, we can easily construct $N$-soliton solutions 
by setting the input data of the $N\times N$ diagonal matrix $\Lambda$ 
and $2\times N$ matrix $\theta$ which satisfy \eqref{initial}. 
For the $N=3$ case, we have 
\begin{eqnarray}
\label{3ss-matrix-pair}
\theta
:=
\left(
\begin{array}{ccc}
a_{1}^2e^{\xi_{1}} & a_{2}^2e^{\xi_{2}} & a_{3}^2e^{\xi_{3}}
\\
b_{1}^2e^{-\xi_{1}} & b_{2}^2e^{-\xi_{2}}& b_{3}^2e^{-\xi_{3}}
\end{array}
\right),~~
\Lambda
:=
\left(
\begin{array}{ccc}
\lambda_{1} & 0 & 0
\\
0 & \lambda_{2} & 0
\\
0 & 0 & \lambda_{3} 
\end{array}
\right), 
\end{eqnarray}
where the parameters are defined as in \eqref{Xi} for $j\in\{1,2,3\}$.

We note that the input data are much simpler than those for quasi-Wronskian solutions \cite{Hamanaka-2023}. 
More specifically, we can construct multi-soliton solutions with arbitrary soliton  numbers without applying the iterations of Darboux transformation.

\section{Multiple-pole Solutions}
As we indicated in the previous section, there is a type of nontrivial solution called the double-pole solution, which could be derived by taking the limit $\lambda_{2} \rightarrow \lambda_{1}$ (with the constraint $\alpha_{1}=\alpha_{2}, \beta_{1}=\beta_{2}$) on the  two-soliton. 
In this limit, both the normal vectors and the amplitudes of the  two-line solitons converge to the same value [cf. \eqref{Xi}, \eqref{Xj}, and \eqref{Asymptotic_2-Soliton}]. Nevertheless, we will see in  Sec. 5.2  
that in the 2D slice [$(w,\widetilde w)=(0,0)$], 
the two solitons converge not to parallel  two-line solitons but to two ``curved solitons'' 
along  with a linear function background  due to resonance effects in the interacting region.  More details will be provided in  Sec. 5.2.  

In this section, we aim to construct the double-pole solutions from the two-soliton solutions \eqref{2ss-matrix-pair} by taking some nontrivial transformations and using L'H$\hat{\mbox{o}}$pital's rule, and then calculate the corresponding  NL$\sigma$M action density.

\subsection{Double-pole Solutions}\label{sec-5-1}
In this section, we show that the double-pole solution can be considered as 
the
$\lambda_2 \rightarrow \lambda_1$ 
limit of the  two-soliton solution \eqref{2ss-matrix-pair}.
For convenience of the later discussion, we introduce the following transformation:
\begin{eqnarray}
\label{Def-new-set}
(\theta^{\prime}, \Lambda^{\prime})=(\theta P^{-1}, P\Lambda P^{-1}),~\Omega^{\prime}(\theta^{\prime}, \theta^{\prime})=P^{-\dagger}\Omega(\theta, \theta) P^{-1}.
\end{eqnarray}
Direct calculation shows that the linear system and the Sylvester equation are form invariant under \eqref{Def-new-set}. More explicitly, we have
\begin{eqnarray}
\left\{
\begin{array}{l}
\left[\partial_{z} - (\partial_{z}J)J^{-1} \right]\theta^{\prime} -(\partial_{\widetilde{w}}\theta^{\prime})\Lambda^{\prime} = 0
\smallskip \\
\left[\partial_{w} - (\partial_{w}J)J^{-1} \right]\theta^{\prime} -(\partial_{\widetilde{z}}\theta^{\prime})\Lambda^{\prime} = 0
\end{array}
\right.
\end{eqnarray}
and
\begin{eqnarray}
(\Lambda^{\prime})^{\dagger}\Omega^{\prime}(\theta^{\prime}, \theta^{\prime}) - \Omega^{\prime}(\theta^{\prime}, \theta^{\prime})\Lambda^{\prime} = (\theta^{\prime})^{\dagger}\theta^{\prime}.
\end{eqnarray}
One iteration of the binary Darboux transformation \eqref{J_unitary} shows
\begin{eqnarray}
\label{J-prime = J}
\hat{J}^{\prime}
=
\left|
\begin{array}{cc}
\!\!\Omega^{\prime}(\theta^{\prime}, \theta^{\prime}) & \!\!(\Lambda^{\prime})^{-\dagger}(\theta^{\prime})^{\dagger}
\\
\!\!\theta^{\prime} & \!\!\fbox{$I$}
\end{array}
\!\!\right|\!J 
=
\left|
\begin{array}{cc}
\!\!P^{\dagger}\Omega(\theta, \theta) P^{-1}  & \!\!P^{-\dagger}\Lambda^{-\dagger}\theta^{\dagger}
\\
\!\!\theta P^{-1} & \!\!\fbox{$I$}
\end{array}
\!\!\right|\!J
=
\left|
\begin{array}{cc}
\!\!\Omega(\theta, \theta) & \!\!\Lambda^{-\dagger}\theta^{\dagger}
\\
\!\!\theta & \!\!\fbox{$I$}
\end{array}
\!\!\right|\!J
= \hat{J}, ~
\end{eqnarray}
that is, $\hat{J}$ is invariant under \eqref{Def-new-set}.
Due to this fact, an equivalent expression of the two-soliton solution can be obtained by choosing 
\begin{eqnarray}
P=
\left(
\begin{array}{cc}
\lambda_{1} - \lambda_{2}  &  0
\\
1 & \lambda_{1} - \lambda_{2}
\end{array}
\right)
\end{eqnarray}
in \eqref{Def-new-set}.
In this case, 
\begin{eqnarray}
\theta^{\prime}
=
\left(
\begin{array}{cc}
\displaystyle{\frac{a_{1}^2e^{\xi_{1}}}{\lambda_{1} - \lambda_{2}} - \frac{a_{2}^2e^{\xi_{2}}}{(\lambda_{1} - \lambda_{2})^2}} & \displaystyle{\frac{a_{2}^2e^{\xi_{2}}}{\lambda_{1} - \lambda_{2}}}
\smallskip \\
\displaystyle{\frac{b_{1}^2e^{-\xi_{1}}}{\lambda_{1} - \lambda_{2}} - \frac{b_{2}^2e^{-\xi_{2}}}{(\lambda_{1} - \lambda_{2})^2}} & \displaystyle{\frac{b_{2}^2e^{-\xi_{2}}}{\lambda_{1} - \lambda_{2}}}
\end{array}
\right),~
\Lambda^{\prime}
=
\left(
\begin{array}{cc}
\lambda_{1} & 0 
\\
1 & \lambda_{2}
\end{array}
\right),
\end{eqnarray}
where $\xi_{j}, j=1,2$ are defined in \eqref{Xi}.

Now, let us show that the double-pole solution can be derived by taking  the limit on $(\theta^{\prime}, \Lambda^{\prime})$. By choosing $a_{1}^2=b_{1}^2=1$, $a_{2}^2=b_{2}^2=\lambda_{1}-\lambda_{2}$ and $\alpha_{1}=\alpha_{2}, \beta_{1}=\beta_{2}$, we have
\begin{eqnarray}
\theta^{\prime}
=
\left(
\begin{array}{cc}
\displaystyle{\frac{e^{\xi_{1}} - e^{\xi_{2}}}{\lambda_{1} - \lambda_{2}}} & e^{\xi_{2}}
\smallskip \\
\displaystyle{\frac{e^{-\xi_{1}} - e^{-\xi_{2}}}{\lambda_{1} - \lambda_{2}}} & e^{-\xi_{2}}
\end{array}
\right).
\end{eqnarray}
By taking the limit $\lambda_{2} \rightarrow \lambda_{1}$ and using  L'H$\hat{\mbox{o}}$pital's rule, we have 
\begin{eqnarray}\label{limit-data}
\lim_{\lambda_{2} \rightarrow \lambda_{1}} \theta^{\prime}
= 
\left(
\begin{array}{cc}
\dot{\xi}_{1}e^{\xi_{1}} & e^{\xi_{1}} 
\\
-\dot{\xi}_{1}e^{-\xi_{1}} & e^{-\xi_{1}}
\end{array}
\right),~
\lim_{\lambda_{2} \rightarrow \lambda_{1}} \Lambda^{\prime}
= 
\left(
\begin{array}{cc}
\lambda_{1} & 0 
\\
1 & \lambda_{1}
\end{array}
\right),
\end{eqnarray}
where  $\dot{\xi}_{1}:=\partial_{\lambda_{1}}\xi_{1}$. 
For simplicity, we omit the lower index and rewrite the input data for the double-pole solution as the following $2 \times 2$ matrix pair $(\theta, \Lambda)$:
\begin{eqnarray}
\label{Double-pole solution}
\theta
:=
\left(
\begin{array}{cc}
\dot{\xi}e^{\xi} & e^{\xi} \\ 
-\dot{\xi}e^{-\xi} & e^{-\xi}
\end{array}
\right),~
\Lambda
:=
\left(
\begin{array}{cc}
\lambda & 0 \\
1 & \lambda
\end{array}
\right),~
\end{eqnarray}
where
\begin{eqnarray}
\xi:=\lambda\alpha z + \beta \widetilde{z} + \lambda\beta w + \alpha \widetilde{w},~
\dot{\xi}:=\partial_{\lambda}\xi = \alpha z + \beta w, ~
\lambda, \alpha, \beta, z, \widetilde{z}, w, \widetilde{w} \in \mathbb{C}.
\end{eqnarray}
In fact, such types of solutions with Jordan block matrices $\Lambda$ are obtained in Appendix C.3 in \cite{LQYZ-SAPM-2022}.

The corresponding NL$\sigma$M action density can be calculated as (see Appendix \ref{Double-pole data} for details)
\begin{eqnarray}
	\label{Double-pole_action density}
	&&\!\!\!\!\!\!{\cal{L}}_\sigma
	=
	-\frac{1}{16\pi}\mbox{Tr}\left[
	(\partial_{m}\hat{J})\hat{J}^{-1}(\partial^{m}\hat{J})\hat{J}^{-1}
	\right]  
	\nonumber \\
	&\!\!\!\!=\!\!\!\!&
	\frac{
		-(\alpha\overline{\beta} - \overline{\alpha}\beta)(\lambda - \overline{\lambda})^3
		\left\{
		\begin{array}{l}
			-(\lambda + \overline{\lambda})^2\left[
			1 + \mbox{cosh}(2X)
			\right]
			\smallskip \\
			+2(\lambda - \overline{\lambda})^2
			|\lambda|^2|\dot{\xi}|^2\mbox{cosh}(2X)
			\smallskip \\
			+(\lambda - \overline{\lambda})\left[
			\begin{array}{l}
				(\lambda + \overline{\lambda})(\lambda \dot{\xi} - \overline{\lambda}~\!\dot{\overline{\xi}})
				+|\lambda|^2(\dot{\xi} - \dot{\overline{\xi}})
			\end{array}
			\right]\mbox{sinh}(2X)
		\end{array}
		\right\}
	}
	{
		4\pi|\lambda|^4\left[
		1+2|\lambda - \overline{\lambda}|^2|\dot{\xi}|^2 + \mbox{cosh}(2X)
		\right]^2
	},~~~~~~~~
\end{eqnarray}
where 
\begin{eqnarray}
\label{X}
	X&\!\!\!\!:=\!\!\!\!& ~\xi + \overline{\xi},  \\
	\xi&\!\!\!\!=\!\!\!\!&
	\lambda\alpha z + \beta \widetilde{z} + \lambda\beta w + \alpha \widetilde{w}
	\nonumber \\
	&\!\!\!\!=\!\!\!\!&
	\frac{1}{\sqrt{2}}\left\{
	(\lambda\alpha + \beta)x^1 + (\lambda\beta-\alpha)x^2 + (\lambda\alpha -\beta)x^3 + (\lambda\beta + \alpha)x^4
	\right\}.
\end{eqnarray}

We remark that in the $\lambda_2 \rightarrow \lambda_1$ 
limit, the Wess-Zumino action density also goes to $0/0$ as is seen in Appendix D.4 in \cite{Hamanaka-2023} [where the numerators in (D.16) $\sim$ (D.19) vanish in the $\lambda_2\rightarrow \lambda_1$ limit; then, $a=0$ and $\mathcal{E}_{12}=\mathcal{E}_{21}=\mathcal{e}_{12}=\widetilde{\mathcal{e}}_{12}= 0$, see also (D.15)]. Hence, there would be the same kind of contribution from the Wess-Zumino action density only in the interacting region. (In the asymptotic region, there is no contribution because ${\cal{L}}_{\scriptsize{\mbox{WZ}}}=0$ in any direction.)

\subsection{Asymptotic Analysis of Double-pole Solutions}

In this section, we analyze the asymptotic behavior of \eqref{Double-pole_action density}.
First of all, we introduce  
\begin{eqnarray}
	\label{Z_pm}
	Z_{\pm}&\!\!\!:=\!\!\!&X \mp \log{|\dot{\xi}|}
	\nonumber \\ 
	&\!\!\!=\!\!\!&(\lambda\alpha + \overline{\lambda}\overline{\alpha})z + (\beta + \overline{\beta})\widetilde{z}
	+ (\lambda\beta + \overline{\lambda}\overline{\beta})w + (\alpha + \overline{\alpha})\widetilde{w} 
	\mp \log|\alpha z + \beta w|, ~~
\end{eqnarray}
and consider the following two asymptotic regions:
\begin{eqnarray}
	\label{asymptotic regions}
	\left\{ 
	\begin{array}{l}
		(1)~R_{+}:~ X \rightarrow +\infty, \log{|\dot{\xi}|}\rightarrow +\infty~~\mbox{such that} ~~Z_{+} ~{\mbox{is finite}} 
		\medskip \\
		(2)~R_{-}: ~ X \rightarrow -\infty, \log{|\dot{\xi}|} \rightarrow +\infty~~ \mbox{such that} ~~Z_{-} ~{\mbox{is finite}} 
	\end{array}
	\right..
\end{eqnarray}

Note that $Z_{\pm}=\mathrm{constant}$ describes the logarithmic curves living on a linear function background.

By dividing a common factor $|\dot{\xi}|^2 e^{2X}$ in both the denominator and numerator of \eqref{Double-pole_action density} and considering case (1) of \eqref{asymptotic regions}, 
since $|\lambda\dot{\xi} - \overline{\lambda}\overline{\dot{\xi}}|=
2\mbox{Im}|\lambda\dot{\xi}| \leq
2|\lambda||\dot{\xi}|$ implies 
$ |\lambda\dot{\xi} - \overline{\lambda}\overline{\dot{\xi}}| =O(|\dot{\xi}|)$,
we find that \eqref{Double-pole_action density} is dominated by the following expression\footnote{We briefly introduce the definition of the big O notation.
	If there exist $x_{0}, c$ such that $|f(x)| \leq c|g(x)|$, for all $x > x_{0}$, then we say $f(x)=O(g(x))$.}: 
\begin{eqnarray}
	16\pi{\cal{L}}_\sigma 
	&\!\!\!=\!\!\!&
	\displaystyle{\frac{A + O(|\dot{\xi}|^{-1})}
		{
			\left[(|\dot{\xi}|^{-1}e^{X} + 4|\lambda - \overline{\lambda}|^2|\dot{\xi}|e^{-X})/2 + O(|\dot{\xi}|^{-1}e^{-X})
			\right]^2
	}},
	\\
	&\!\!\! = \!\!\!& 
	\frac{d_{11} + O(|\dot{\xi}|^{-1})}
	{\mbox{cosh}^2\left[X - \log|\dot{\xi}| - \log(2|\lambda - \overline{\lambda}|)\right] + O(|\dot{\xi}|^{-2})},	
	\\
	&\!\!\! \longrightarrow \!\!\!& d_{11}\mbox{sech}^2\left[ X - \log|\dot{\xi}| - \log(2|\lambda - \overline{\lambda}|)\right] 
	=
	d_{11}\mbox{sech}^2( Z_{+} - \delta) ~\mbox{on $R_{+}$}, ~~~~
	\label{Asymptotic_DP_infty}
\end{eqnarray}
where $A:=-16(\alpha\overline{\beta} - \overline{\alpha}\beta)(\lambda - \overline{\lambda})^5/ |\lambda|^2$, $d_{11}$ is defined in \eqref{d_jk} with $\lambda_{1}=\lambda$, and the  phase-shift factor is
$\delta := \log(2|\lambda - \overline{\lambda}|)$.

\begin{figure}[ht]
	\centering
	\subfigure[]{
		\includegraphics[width=0.28\textwidth]{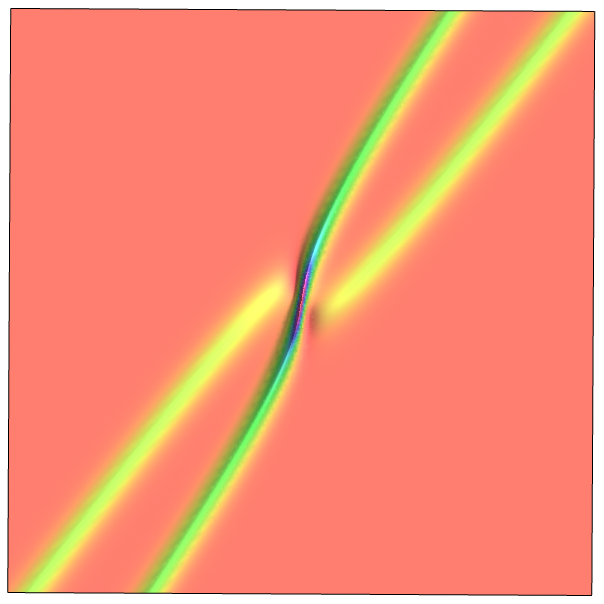}}
	\hspace{0.3cm} 
	\subfigure[]{
		\includegraphics[width=0.3\textwidth]{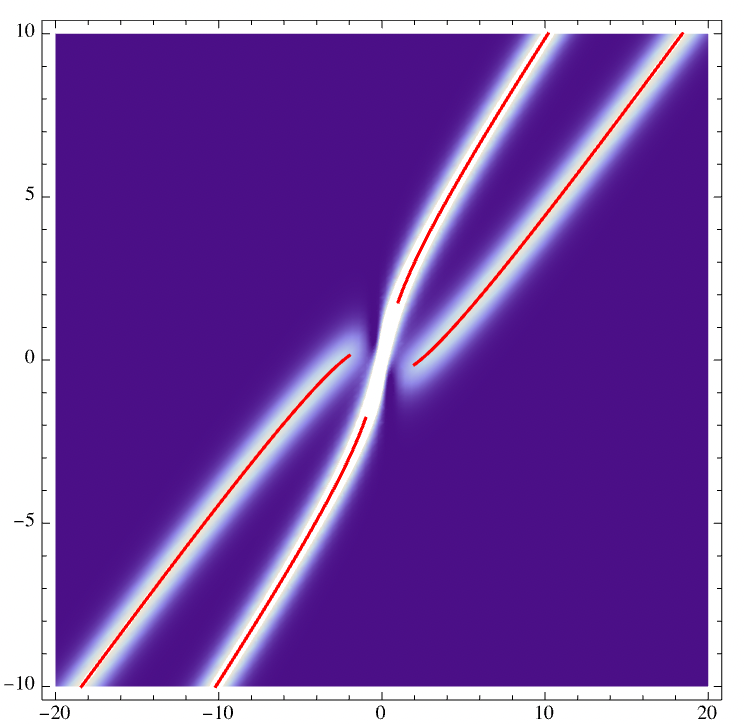}}
	\caption{Plots of the 2D slice of the double-pole soliton NL$\sigma$M action density with $\lambda_1=-1+i,\alpha=0.5-i,\beta=-0.7-1.4i$, $(z,\tilde z)\in[-20,20]\times[-10,10]$, and $(w,\tilde w)=(0,0)$. (a) Shape of the 2D slice of NL$\sigma$M action density. (b) Density plot with four red curves of $Z_{\pm} \mp \delta=0.$
	}
	\label{fig-8}
\end{figure}

Similarly, by dividing another common factor $ |\dot{\xi}|^2e^{-2X}$ in both denominator and numerator of \eqref{Double-pole_action density} and consider case (2) of \eqref{asymptotic regions}, 
we obtain
\begin{eqnarray}
	16\pi{\cal{L}}_\sigma 
	&\!\!\!=\!\!\!&
	\displaystyle{\frac{A + O(|\dot{\xi}|^{-1})}
		{
			\left[(4|\lambda - \overline{\lambda}|^2|\dot{\xi}|e^{X} + |\dot{\xi}|^{-1}e^{-X})/2 + O(|\dot{\xi}|^{-1}e^{X})
			\right]^2
	}},
	\\
	&\!\!\! = \!\!\!& 
	\frac{d_{11} + O(|\dot{\xi}|^{-1})}
	{\mbox{cosh}^2\left[X + \log|\dot{\xi}| + \log(2|\lambda - \overline{\lambda}|)\right] + O(|\dot{\xi}|^{-2})},	
	\\
	&\!\!\! \longrightarrow \!\!\!& d_{11}\mbox{sech}^2\left[X + \log|\dot{\xi}| + \log(2|\lambda - \overline{\lambda}|)\right]
	= d_{11}\mbox{sech}^2(Z_{-} + \delta) ~ \mbox{on $R_{-}$}. ~~~~ \label{Asymptotic_DP_-infty}
\end{eqnarray}
Note that \eqref{Asymptotic_DP_infty} and \eqref{Asymptotic_DP_-infty} involve a logarithm function term, which is quite different from \eqref{Asymptotic_2-Soliton}, although \eqref{Asymptotic_DP_infty}, \eqref{Asymptotic_DP_-infty}, and \eqref{Asymptotic_2-Soliton} are quite similar in form at first glance.
Therefore, we find that the two peaks of action density in \eqref{Double-pole_action density}  are localized on 
two curved  hypersurfaces $Z_{\pm} \mp \delta=0$.
This is illustrated by taking a 2D slice $(w, \widetilde{w})=(0, 0)$ 
in Fig. \ref{fig-8}, where
$Z_{\pm}=(\lambda\alpha + \overline{\lambda}\overline{\alpha})z + (\beta + \overline{\beta})\widetilde{z} \mp \log|\alpha| \mp \log|z|$. 

\subsection{Multiple-pole Solutions}\label{sec-5-2}

In a similar way, we can construct multiple-pole (of degree $N$) solutions  
by setting the input data of an $N\times N$ Jordan block matrix $\Lambda$  (instead of a diagonal matrix) 
For such solutions, the input data are given by
\begin{align}
	\theta=
	\begin{pmatrix}
		\frac{\partial_\lambda^{N-1}(e^\xi)}{(N-1)!} & \cdots & \partial_\lambda(e^\xi) & e^\xi \\
		\frac{\partial_\lambda^{N-1}(e^{-\xi})}{(N-1)!} & \cdots & \partial_\lambda(e^{-\xi}) & e^{-\xi} \\
	\end{pmatrix},~
	\Lambda=
	\begin{pmatrix}
		\lambda & 0 & \cdots & 0 \\
		1 & \lambda & \cdots & 0 \\
		\vdots & \vdots & \ddots & 0 \\
		0 & 0 & \cdots & \lambda
	\end{pmatrix}.
\end{align}

For example, if we take $N=3$, the input data for the triple-pole solution will be
\begin{eqnarray}
\label{triple-pole solution}
\theta=
\left(
\begin{array}{ccc}
\frac{1}{2}[\ddot{\xi}+(\dot\xi)^2]e^{\xi}&\dot{\xi}e^{\xi} & e^{\xi} \\ 
\frac{1}{2}[-\ddot{\xi}+(\dot\xi)^2]e^{-\xi} &-\dot{\xi}e^{-\xi} & e^{-\xi}
\end{array}
\right),~
\Lambda=
\left(
\begin{array}{ccc}
\lambda & 0  & 0 \\
1 & \lambda & 0 \\
0 & 1 & \lambda 
\end{array}
\right),
\end{eqnarray}
where $\ddot{\xi}:=\partial^2_\lambda\xi$, and the parameters are defined as in \eqref{Xi} and \eqref{X}. 
This can be realized as the 
$\lambda_2 \rightarrow \lambda_1$ and $\lambda_3 \rightarrow \lambda_1$ 
limits [cf. Eq. \eqref{3ss-matrix-pair}] of three-line solitons with equal amplitude. 


\section{Examples and Figures}

In Secs. \ref{sec-4-1} and \ref{sec-5-1}, the NL$\sigma$M action densities of the two-soliton solution and double-pole solution are shown. Their connection via a limiting procedure is revealed in  Sec. \ref{sec-4-2}. 
In this section, we explain the resonance limits of two-soliton solutions by illustrating fruitful figures in terms of the explicit result \eqref{NL Sigma term_2-Soliton_form 2}, where the double-pole interaction \eqref{Double-pole_action density} and V-shape soliton will be revealed.

Since $\cal L_{\sigma}$ is a four-dimensional function of $(z,\tilde z,w,\tilde w)$, to illustrate the shape, we can draw a two-dimensional slice of it with flexible $(z,\tilde z)$ and fixed $(w,\tilde w)$. For convenience, we will choose $(w,\tilde w)=(0,0)$ in later illustrations.

\subsection{One-Soliton solution}
\begin{figure}[ht]
	\centering
	\subfigure[]{
		\includegraphics[width=0.4\textwidth]{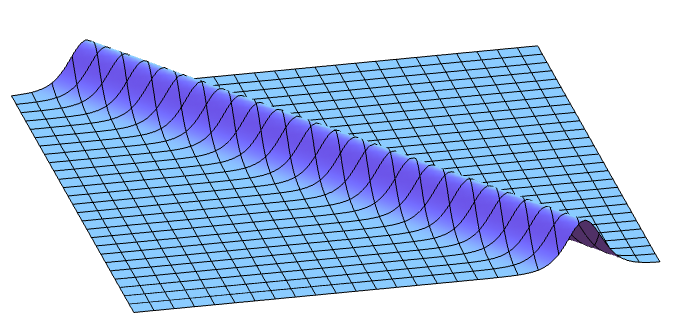}}
	\hspace{0.5cm} 
	\subfigure[]{
		\includegraphics[width=0.22\textwidth]{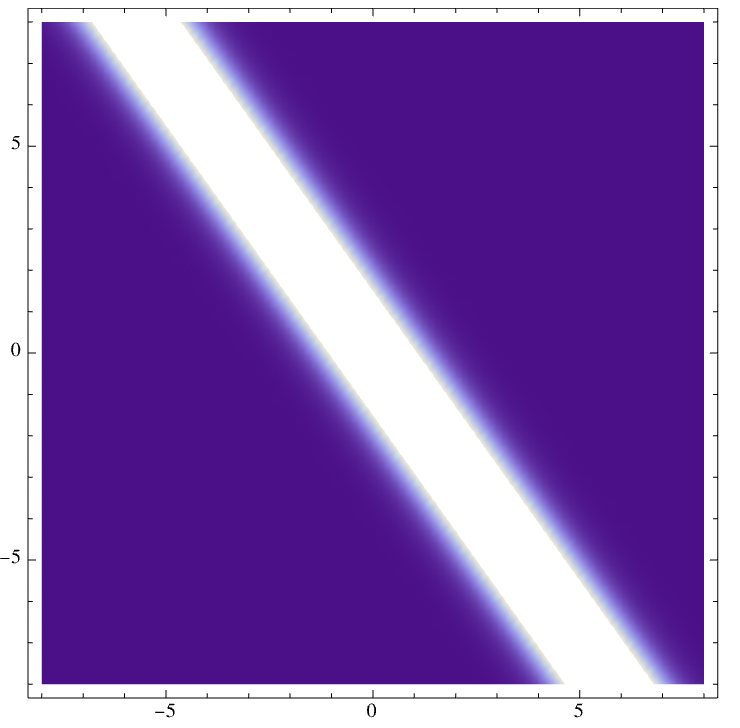}}
	\caption{Plots of the 2D slice of two-soliton NL$\sigma$M action density with $\lambda_1=0.5+0.5i,\alpha_1=0.5-0.5i,\beta_1=-0.7-1.4i$, $(z,\tilde z)\in[-8,8]\times[-8,8]$, and $(w,\tilde w)=(0,0)$. (a) Shape of the 2D slice of NL$\sigma$M action density. (b) Density plot of the 2D slice of NL$\sigma$M action density.}
	\label{fig-1SS}
\end{figure}

We start with the one-soliton case, and the corresponding NL$\sigma$M action density is given by \eqref{NLsigmaM-1SS}. The shape and density plot of the two-dimensional slice are illustrated in  Fig. \ref{fig-1SS}.

\subsection{Two-Soliton Solution}
\begin{figure}[ht]
	\centering
	\subfigure[]{
		\includegraphics[width=0.4\textwidth]{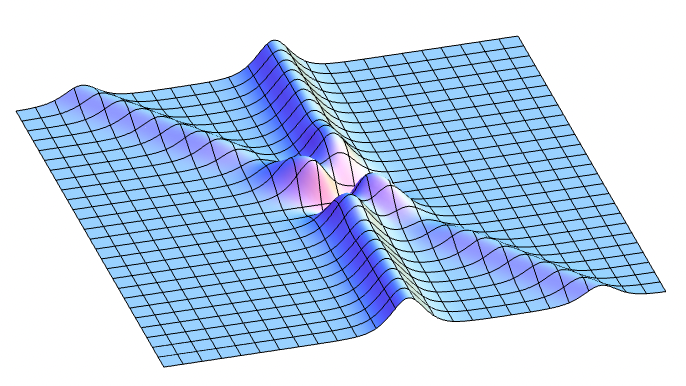}}
	\hspace{0.5cm} 
	\subfigure[]{
		\includegraphics[width=0.22\textwidth]{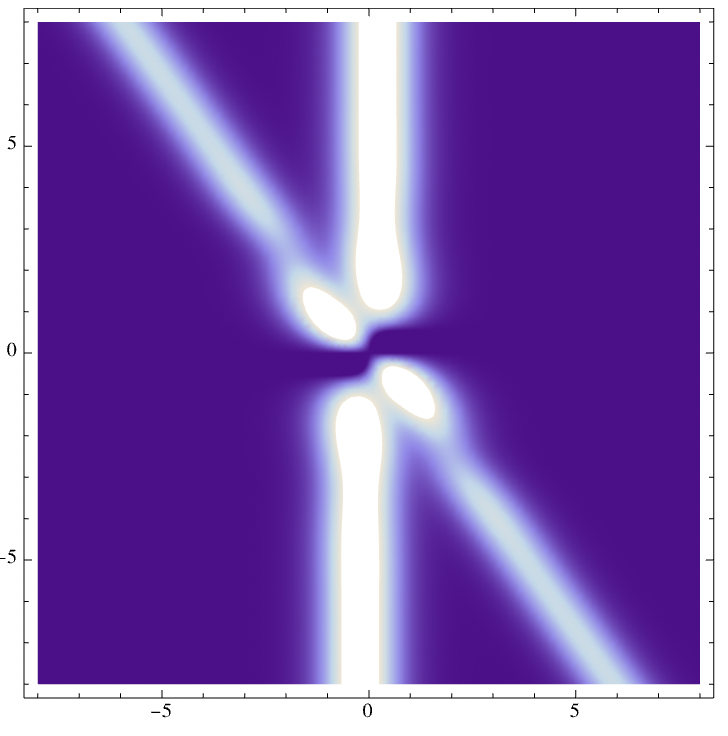}}
	\caption{Plots of the 2D slice of two-soliton NL$\sigma$M action density with $\lambda_1=-1+i,\lambda_2=0.5+0.5i,\alpha_1=\alpha_2=0.5-0.5i,\beta_1=\beta_2=-0.7-1.4i$, $(z,\tilde z)\in[-8,8]\times[-8,8]$, and $(w,\tilde w)=(0,0)$. (a) Shape of the 2D slice of NL$\sigma$M action density. (b) Density plot of the 2D slice of NL$\sigma$M action density.}
	\label{fig-1}
\end{figure}

For two-soliton solutions, the shape and density plots of the two-dimensional slice of NL$\sigma$M action density are illustrated in Fig. \ref{fig-1}, where we can clearly see there is a gap in the interacting region. This fact coincides with our conjecture that the two-soliton may be decomposed to two V-shape solitons by taking large phase-shift limits.

\subsection{Double-Pole Solution}
\begin{figure}[ht]
	\centering
	\subfigure[]{
			\includegraphics[width=0.4\textwidth]{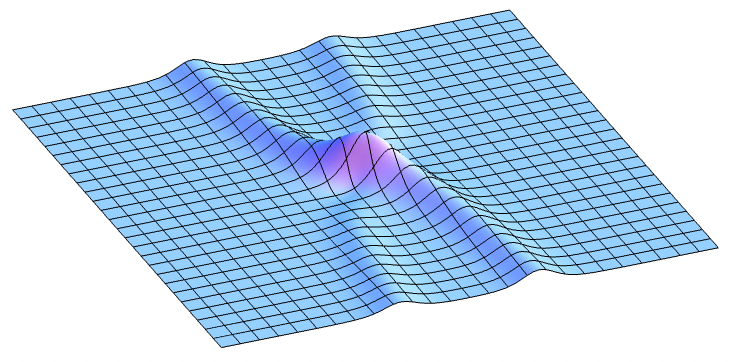}}
	\hspace{0.5cm} 
	\subfigure[]{
			\includegraphics[width=0.22\textwidth]{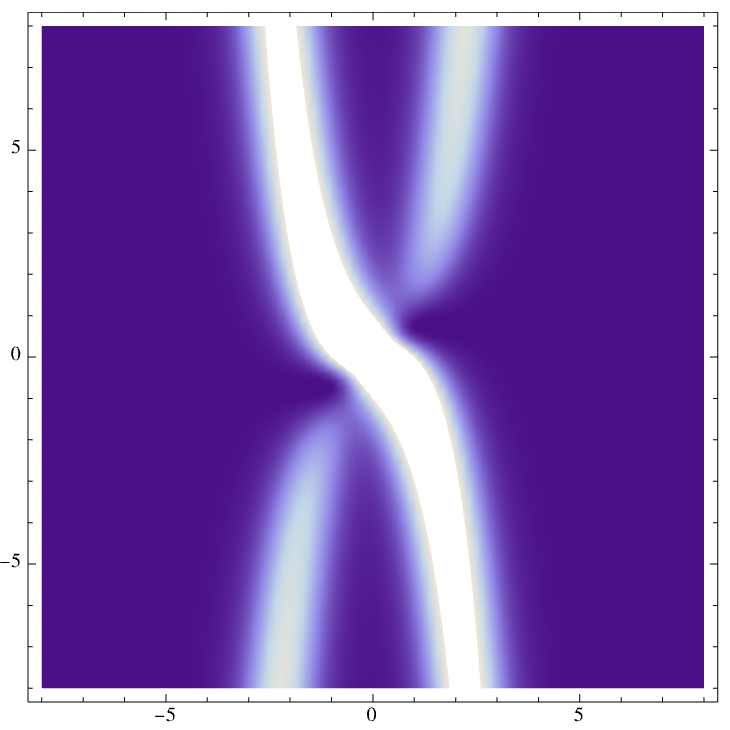}}
	\caption{Plots of the 2D slice of double-pole soliton NL$\sigma$M action density with $\lambda_1=-1+i,\alpha=0.5-0.5i,\beta=-0.7-1.4i$, $(z,\tilde z)\in[-8,8]\times[-8,8]$, and $(w,\tilde w)=(0,0)$. (a) Shape of the 2D slice of NL$\sigma$M action density. (b) Density plot of the 2D slice of NL$\sigma$M action density.}
	\label{fig-2}
\end{figure}

Now, we consider the case of double-pole solutions. According to the result in \eqref{Double-pole_action density}, the shape and density plots of the two-dimensional slice of NL$\sigma$M action density are illustrated in Fig. \ref{fig-2}. By taking the 
$\lambda_2 \rightarrow \lambda_1$ 
limit of two-line solitons 
together with $\alpha_1=\alpha_2$ and $\beta_1=\beta_2$, 
the gap in the two-soliton interaction vanishes and becomes a hill.
Note that in general, in double-pole solution, 
the branches are asymptotically governed by the logarithmic function of $z$ (e.g., see \cite{Zhang-2014}). 
This is true of our case as is seen in Sec. 5.2.

\begin{figure}[ht]
	\centering
	\subfigure[]{
		\includegraphics[width=0.22\textwidth]{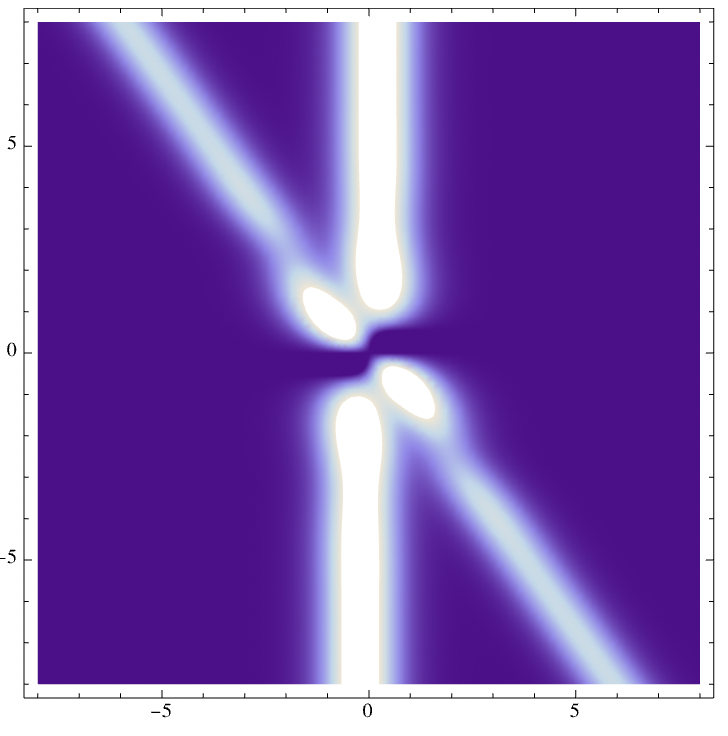}}
	\hspace{0.2cm} 
	\subfigure[]{
		\includegraphics[width=0.22\textwidth]{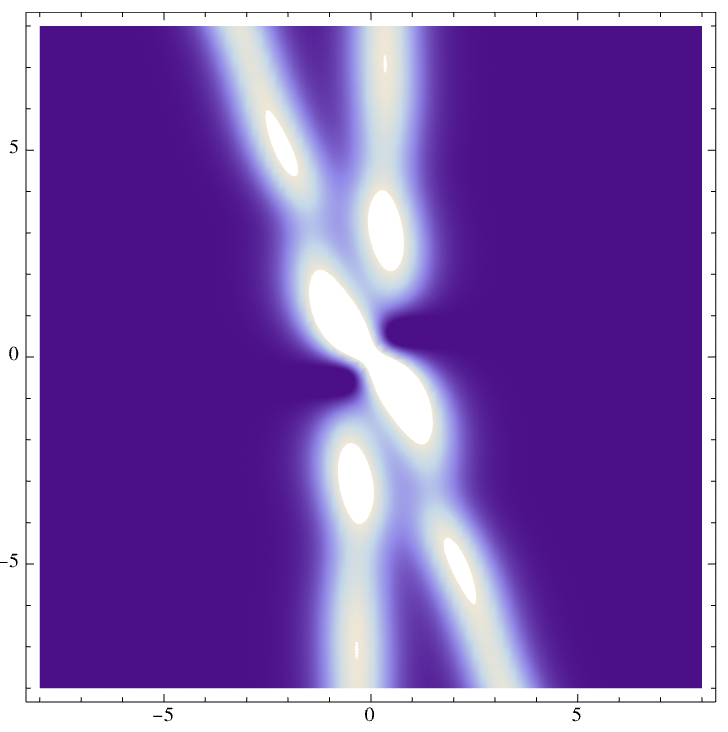}}
	\subfigure[]{
		\includegraphics[width=0.22\textwidth]{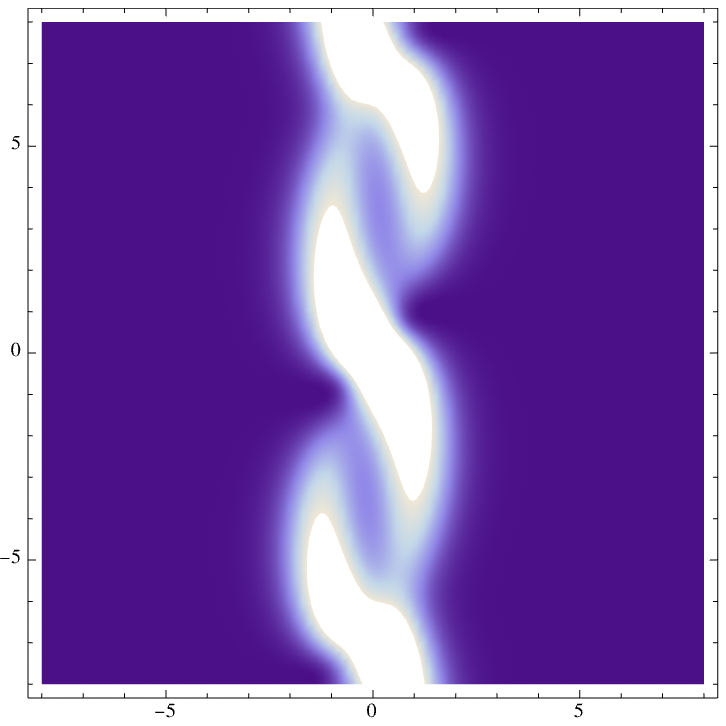}}
	\hspace{0.2cm} 
	\subfigure[]{
		\includegraphics[width=0.22\textwidth]{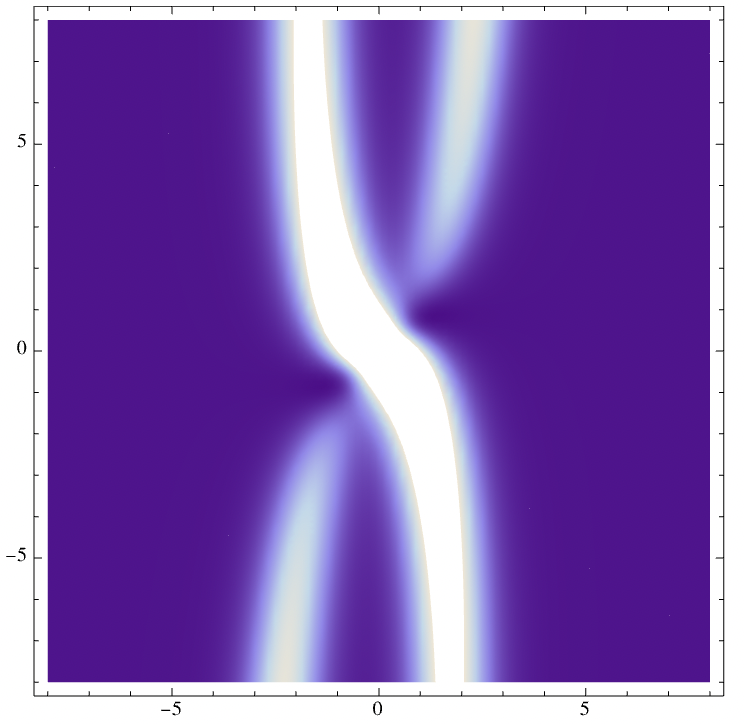}}
	\caption{Density plots of the 2D slice of two-soliton NL$\sigma$M action density with $\alpha_1=\alpha_2=0.5-0.5i,\beta_1=\beta_2=-0.7-1.4i$, $(z,\tilde z)\in[-8,8]\times[-8,8]$ and $(w,\tilde w)=(0,0)$. (a) Plot with $\lambda_1=-1+i$ and $\lambda_2=0.5+0.5i$. (b) Plot with  $\lambda_1=-1+i$ and $\lambda_2=0.5i$. (c) Plot with $\lambda_1=-1+i$ and $\lambda_2=-0.6+0.5i$. (d) Plot with $\lambda_1=-1+i$ and $\lambda_2=-0.9+0.8i$.}
	\label{fig-3}
\end{figure}

In  Sec. \ref{sec-5-1}, we claimed that the double-pole solution can be regarded as a limit of the two-soliton solution by choosing parameters \eqref{limit-data}. On the other hand, this fact can be verified by illustrating various two-soliton action density figures with $\lambda_2$ going to $\lambda_1$. In  Fig. \ref{fig-3}, we illustrate different plots of the two-soliton, with $|\lambda_2-\lambda_1|$ becoming extremely small. This result coincides with the plot in  Fig. \ref{fig-2}.

\begin{figure}[ht]
	\centering
	\subfigure[]{
		\includegraphics[width=0.4\textwidth]{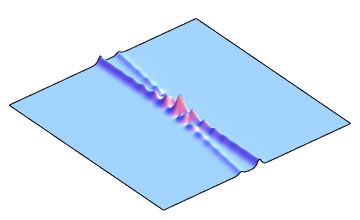}}
	\hspace{0.5cm} 
	\subfigure[]{
		\includegraphics[width=0.25\textwidth]{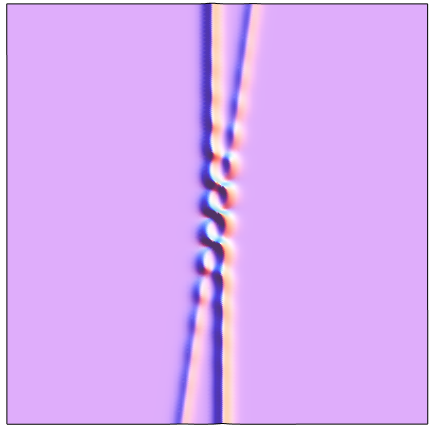}}
	\caption{Shape of the two soliton interaction in  Fig.} \ref{fig-3}(c), with $(z,\tilde z)\in[-50,50]\times[-25,25]$.
	\label{fig-5}
\end{figure}

Notice that in Fig. \ref{fig-3}(c) [see also in  Fig. \ref{fig-5}, as Fig \ref{fig-3}(c) is a zoom in of Fig. \ref{fig-5}], the two soliton interaction behaves like a DNA double helix structure, which exhibits periodical behavior in the interaction area. 

\subsection{V-Shape Solution}

As we mentioned in Sec. \ref{sec-4-1}, the V-shape solution appears when we take the large  phase-shift limit of a two-soliton. 
From \eqref{Phase_Shift_Factor}, the  phase-shift factor $\widetilde\delta$ grows when we take the limit $\lambda_2\rightarrow\overline\lambda_1$.
In the case (2) of \eqref{Reduced solutions of 2-soliton}, the NL$\sigma$M action density $\cal{L}_\sigma$ becomes zero when $\lambda_2\rightarrow\overline\lambda_1$ and, during this procedure, the gap in the two-soliton interaction grows larger, hence two V-shape solitons will emerge. 

\section{Conclusion and Discussion}

In this paper, we discovered resonance soliton solutions in the WZW$_4$ model in Ultrahyperbolic space $\mathbb{U}$ 
from the perspective of the CMA and the binary Darboux transformation. 
We found that the exact solutions in \cite{LQYZ-SAPM-2022} can be rewritten as the quasi-Grammian form, so that some quasideterminant techniques can be applied to investigate various relations between the CMA \cite{LQYZ-SAPM-2022} and binary Darboux transformation. According to our present research, the  CMA appears to be a particular class of the binary Darboux transformation in \cite{Nimmo-2000}, and we established a generalized CMA which admits simple input data for the soliton solutions. For the one-soliton and two-soliton cases, we found that the quasi-Grammian and quasi-Wronskian soliton solutions gave the same action densities. It would be worthwhile to investigate whether this is a general result for $n$-solitons.


One highlight of this paper is that we found new classical solutions in the WZW$_4$ model by observing resonance limits from two-soliton solutions. One resonance solution was the double-pole solution. The NL$\sigma$M action density of this solution \eqref{Double-pole_action density} revealed a new nonlinear phenomenon different from the two-soliton case \eqref{NL Sigma term_2-Soliton_form 2}. Furthermore, by considering the large  phase-shift limit of two-solitons, we discovered V-shape solitons in pairs. This suggests pair annihilations or creations of  two-line solitons in the open $N=2$ string theory. 
 
In the classification of the KP line solitons, Y-shape resonance solitons are building blocks of the soliton interaction diagrams and are essential. Likewise, we can expect that the V-shape solitons would be building blocks to classify the ASDYM solitons including the resonance processes. Furthermore,  V-shape solitons also exist in the CBS equation \cite{IK-2001} and Zakharov systems \cite{IKT-2002}, both of which can be derived from the ASDYM equation by dimensional reduction  \cite{Schiff-1992, Strachan-1992-N}.   Therefore, the hidden symmetry behind the ASDYM equation would be described by the 2-toroidal algebra \cite{NS-1992,Losev-1996,Inami-1997-T} because the toroidal algebras are the hidden symmetry behind the CBS equation \cite{IK-2001} and Zakharov systems \cite{IKT-2002}.


On the other hand, the hidden symmetry behind the KP equations is described by W$_{1+\infty}$ algebras  \cite{Sato-1983}--\cite{Yu-1991}. 
There seems to be no solutions of either V-shape solitons or smooth multiple-pole solitons for the KP equations (exactly speaking the KPII equations.\footnote{There are two kinds of KP equations: the KPI and KPII equations. The latter describe shallow water waves and therefore are popular in the study of integrable systems. In this paper, we refer to the KPII equations only. For the KPI equation, there exists smooth multiple-pole solutions \cite{AV-1999}.}) 
The KP equations have not yet been derived from the ASDYM equations by reduction.\footnote{If we allow  operator-valued} gauge fields, the KP equation can be reduced from the ASDYM equation \cite{ACT-1993}. However, in this case, it lacks any twistor descriptions and, hence, is not considered as an example of the Ward conjecture.  
Therefore, our present results would suggest new aspects different from the KP solitons and lead to a new classification theory of the ASDYM solitons.


\subsection*{Acknowledgements}

MH thanks the organizers, especially Frank Nijhoff, for their invitation, hospitality, and discussion at the BIRS-IASM workshop on Lagrangian Multiform Theory and Pluri-Lagrangian Systems, Hangzhou, China, in October 2023, and Shanghai University (and Da-jun Zhang) for their support and hospitality during the stay from 24 December 2023 to 11 January 2024, where this collaboration started. MH is also grateful to Yuji Kodama, Kenichi Maruno, and Kanehisa Takasaki for their useful comments. 
The work of MH is supported in part by the Ichihara International Foundation, the Chubei Ito Foundation, and the Daiko Foundation. 
The work of SCH is supported in part by 
the Iwanami Fujukai Foundation and the Daiko Foundation. 
DJZ is supported by the China-Japan Scientific Cooperation Program between NSFC and JSPS (No. 12411540016), the NSFC grant (No. 12271334), and the Senior Research Fellow grant from China Scholarship Council (No. 202406890074).
The work of SSL is supported by China Scholarship Council (No. 202306890061).

\begin{appendices}

\section{Proof of Lemma 3.1}
For the Sylvester equation [Eq. \eqref{Sylvester eq}] and the differential recurrence \eqref{Differential recurrence_x_j}, we have
\begin{eqnarray}
\label{Sylvester eq_derivative_j}
K(\partial_{x_{j}}M)-(\partial_{x_{j}}M)L
=(\partial_{x_{j}}r)s^{T} + r(\partial_{x_{j}}s^{T})
\end{eqnarray}
and
\begin{eqnarray}
\label{Sylvester equ_derivative_j+1}
K(\partial_{x_{j+1}}M) - (\partial_{x_{j+1}}M)L
=(\partial_{x_{j+1}}r)s^{T} + r(\partial_{x_{j+1}}s^{T})
=K(\partial_{x_{j}}r)s^{T} + r(\partial_{x_{j}}s^{T})L.
\end{eqnarray}
By calculating $K \times \eqref{Sylvester eq_derivative_j} + \eqref{Sylvester eq_derivative_j} \times L$ and using \eqref{Sylvester equ_derivative_j+1}, one obtains
\begin{eqnarray}
K^{2}(\partial_{x_{j}}M) - (\partial_{x_{j}}M)L^{2}
&\!\!\!\!=\!\!\!\!&
K(\partial_{x_{j}}r)s^{T} + Kr(\partial_{x_{j}}s^{T}) +
(\partial_{x_{j}}r)s^{T}L + r(\partial_{x_{j}}s^{T})L 
\nonumber \\
&\!\!\!\!=\!\!\!\!&
K(\partial_{x_{j+1}}M)-(\partial_{x_{j+1}}M)L + (\partial_{x_{j}}r)s^{T}L + Kr(\partial_{x_{j}}s^{T})
\end{eqnarray}
which implies 
\begin{eqnarray}
K[(\partial_{x_{j+1}}M) + r(\partial_{x_{j}}s^{T}) - K(\partial_{x_{j}}M)]
-
[(\partial_{x_{j+1}}M) - (\partial_{x_{j}}r)s^{T} - (\partial_{x_{j}}M)L]L = 0.
\end{eqnarray}
Let us define a new matrix $C$ as:
\begin{eqnarray}
C:=\partial_{x_{j+1}}M + r(\partial_{x_{j}}s^{T}) - K(\partial_{x_{j}}M)
\stackrel{\eqref{Sylvester eq_derivative_j}}
{=}\partial_{x_{j+1}}M - (\partial_{x_{j}}r)s^{T} - (\partial_{x_{j}}M)L
\end{eqnarray}
For the Sylvester equation $KC-CL=0$, where $K$ and $L$ do not have the same eigenvalues, $C=0$ is the unique solution. 
Thus, 
\begin{eqnarray}
\partial_{x_{j+1}}M
=(\partial_{x_j}r)s^{T} + (\partial_{x_j}M)L
=K(\partial_{x_j}M) - r(\partial_{x_j}s^{T}).
\end{eqnarray}
$\hfill\Box$ \\

\section{Calculation of the Two-Soliton }
\label{Calculation of Two-Soliton}
Substituting \eqref{2ss-matrix-pair} into the Sylvester equation [Eq. \eqref{Sylvester eq_hermitian}], we can obtain the entries of $\Omega:=(\Omega_{jk})$ as
\begin{eqnarray}
\label{Omega_jk_2ss}
\Omega_{jk}
=
\frac{(\overline{a}_{j}a_{k})^2e^{\overline{\xi}_{j}+ \xi_{k}} + (\overline{b}_{j}b_{k})^2e^{-(\overline{\xi}_{j}+ \xi_{k})}}{\overline{\lambda}_{j} - \lambda_{k}}, ~j, k =1,2.
\end{eqnarray}
The determinant of $\Omega$ is
\begin{eqnarray}
\label{Omega_det_2ss}
|\Omega|
=
\frac{
	\left\{
	\begin{array}{l}
	~~c_{1}\left[
	|a_{1}|^4|a_{2}|^4e^{X_{1} + X_{2}} + |b_{1}|^4|b_{2}|^4e^{-(X_{1} + X_{2})}
	\right]
	\smallskip \\
	+
	c_{2}\left[
	|a_{1}|^4|b_{2}|^4e^{X_{1} - X_{2}} + |a_{2}|^4|b_{1}|^4e^{-(X_{1} - X_{2})}
	\right]
	\smallskip \\
	+
	c_{3}\left[
	(a_{1}\overline{a}_{2}\overline{b}_{1}b_{2})^2e^{i\Theta_{12}} + (\overline{a}_{1}a_{2}b_{1}\overline{b}_2)^2e^{-i\Theta_{12}}
	\right]
	\end{array}
	\right\}
}
{c_{2}c_{3}},
\end{eqnarray} where
\begin{subequations}
\begin{eqnarray}
c_{1}&\!\!\!\!:=\!\!\!\!&(\lambda_{1}-\lambda_{2})(\overline{\lambda}_{1}-\overline{\lambda}_{2}),~
c_{2}:=(\lambda_{1}-\overline{\lambda}_{2})(\overline{\lambda}_{1}-\lambda_{2}),~
c_{3}:=(\lambda_{1}-\overline{\lambda}_{1})(\lambda_{2}-\overline{\lambda}_{2}),
\\
X_{j}&\!\!\!\!:=\!\!\!\!&\xi_{j} + \overline{\xi}_{j},~
\Theta_{j}:=-i(\xi_{j}-\overline{\xi}_{j}),~ \Theta_{jk}:=\Theta_{j} - \Theta_{k}.
\end{eqnarray}
\end{subequations}
Substituting \eqref{2ss-matrix-pair} and \eqref{Omega_jk_2ss} into \eqref{J_unitary} and then applying \eqref{Quasideteminant_defn_2} and \eqref{Omega_det_2ss}, we have
\begin{eqnarray}
\hat{J}:=
\frac{1}{\Delta}
\left(
\begin{array}{cc}
\Delta_{11} & \Delta_{12} 
\\
\Delta_{21} & \Delta_{22}
\end{array}
\right),
\end{eqnarray}
where
\begin{eqnarray}
\begin{array}{l}
\Delta
:= 
\overline{\lambda_{1}}\overline{\lambda_{2}}c_{2}c_{3}|\Omega|,
\medskip \\
\Delta_{11}
:=
\left\{
\begin{array}{l}
~~c_{1}\left[
\lambda_{1}\lambda_{2}|a_{1}|^4|a_{2}|^4e^{X_{1} + X_{2}} + \overline{\lambda}_{1}\overline{\lambda}_{2}|b_{1}|^4|b_{2}|^4e^{-(X_{1} + X_{2})}
\right]
\smallskip \\
+
c_{2}\left[
\lambda_{1}\overline{\lambda}_{2}|a_{1}|^4|b_{2}|^4e^{X_{1} - X_{2}} + \overline{\lambda}_{1}\lambda_{2}|a_{2}|^4|b_{1}|^4e^{-(X_{1} - X_{2})}
\right]
\smallskip \\
+
c_{3}\left[
|\lambda_{1}|^2(a_{1}\overline{a}_{2}\overline{b}_{1}b_{2})^2e^{i\Theta_{12}} 
+ |\lambda_{2}|^2(\overline{a}_{1}a_{2}b_{1}\overline{b}_2)^2e^{-i\Theta_{12}}
\right]
\end{array}
\right\}, ~~\Delta_{22}= \overline{\Delta}_{11},
\medskip \\
\Delta_{12}
:=
\left\{
\begin{array}{l}
~~(a_{2}\overline{b}_{2})^2
\left[
c_{4}|a_{1}|^4e^{X_{1} + i\Theta_{2}} 
- 
\overline{c}_{4}|b_{1}|^4e^{-X_{1} + i\Theta_{2}}
\right]
\smallskip \\
+
(a_{1}\overline{b}_{1})^2
\left[
c_{5}|a_{2}|^4e^{X_{2} + i\Theta_{1}} 
- 
\overline{c}_{5}|b_{2}|^4e^{-X_{2} + i\Theta_{1}}
\right]
\end{array}
\right\}, ~~\Delta_{21}:=-\overline{\Delta}_{12},
\medskip \\
~~~~~~~~~~~c_{4}:=\overline{\lambda}_{1}(\lambda_{1}-\lambda_{2})(\lambda_{1} - \overline{\lambda}_{2})(\lambda_{2} - \overline{\lambda}_{2}),~
c_{5}:=\overline{\lambda}_{2}(\lambda_{1}-\lambda_{2})(\lambda_{1} - \overline{\lambda}_{1})(\overline{\lambda}_{1} - \lambda_{2}).
\end{array}
\end{eqnarray}
The NL$\sigma$M action density is
\begin{eqnarray}
{\cal{L}}_\sigma&=&
-\frac{1}{16\pi}
\mbox{Tr}\left[(\partial_{\mu}\hat{J})\hat{J}^{-1}(\partial^{\mu}\hat{J})\hat{J}^{-1}\right] \nonumber  \\
&\!\!\!\! = \!\!\!\!&
\label{NL Sigma term_2-Soliton_form 1}
\frac{
	\left\{ 
	\begin{array}{l}
	~c_{1}c_{2}
	\displaystyle{\left[
		d_{11}\left(
		\frac{|a_{2}|^2}{|b_{2}|^2}e^{X_{2}}
		+
		\frac{|b_{2}|^2}{|a_{2}|^2}e^{-X_{2}}
		\right)^2
		+ d_{22}\left(
		\frac{|a_{1}|^2}{|b_{1}|^2}e^{X_{1}}
		+
		\frac{|b_{1}|^2}{|a_{1}|^2}e^{-X_{1}}
		\right)^2
		\right]
	}
	\medskip \\
	\!\!+ c_{1}c_{3}
	\left[
	\begin{array}{l}
	~~\displaystyle{d_{12}\left(
		\frac{\overline{a}_{1}a_{2}}{\overline{b}_{1}b_{2}}e^{\frac{X_1 + X_2 - i\Theta_{12}}{2}}
		+
		\frac{\overline{b}_{1}b_{2}}{\overline{a}_{1}a_{2}}e^{-(\frac{X_1 + X_2 - i\Theta_{12}}{2})}
		\right)^2
	}
	\smallskip \\
	+ \displaystyle{\overline{d}_{12}\left(
		\frac{a_{1}\overline{a}_{2}}{b_{1}\overline{b}_{1}}e^{\frac{X_1 + X_2 + i\Theta_{12}}{2}}
		+\frac{b_{1}\overline{b}_{2}}{a_{1}\overline{a}_{2}}e^{-(\frac{X_1 + X_2 + i\Theta_{12}}{2})}
		\right)^2
	}
	\end{array}
	\right]
	\medskip \\
	\!\!- c_{2}c_{3}\left[
	\begin{array}{l}
	~~\displaystyle{f_{12}~\!\left(
		\frac{\overline{a}_{1}\overline{b}_{2}}{\overline{a}_{2}\overline{b}_{1}}e^{\frac{X_1 - X_2 - i\Theta_{12}}{2}}
		-\frac{\overline{a}_{2}\overline{b}_{1}}{\overline{a}_{1}\overline{b}_{2}}e^{-(\frac{X_1 - X_2 - i\Theta_{12}}{2})}
		\right)^2
	}
	\smallskip \\
	+\displaystyle{\overline{f}_{12}~\!\left(
		\frac{a_{1}b_{2}}{a_{2}b_{1}}e^{\frac{X_1 - X_2 + i\Theta_{12}}{2}}
		-\frac{a_{2}b_{1}}{a_{1}b_{2}}e^{-(\frac{X_1 - X_2 +i\Theta_{12}}{2})}
		\right)^2
	}
	\end{array}
	\right]
	\end{array}
	\!\!\!\right\}
}
{2\pi\displaystyle{
		\left[
		\begin{array}{l}
		~~c_{1}\displaystyle{
			\left(
			\frac{|a_{1}|^2|a_{2}|^2}{|b_{1}|^2|b_{2}|^2}e^{X_1 + X_2}   
			+\frac{|b_{1}|^2|b_{2}|^2}{|a_{1}|^2|a_{2}|^2}e^{-(X_1 + X_2)} 
			\right)
		}
		\smallskip \\
		+ c_{2}\displaystyle{
			\left(
			\frac{|a_{1}|^2|b_{2}|^2}{|a_{2}|^2|b_{1}|^2}e^{X_1 - X_2}   
			+\frac{|a_{2}|^2|b_{1}|^2}{|a_{1}|^2|b_{2}|^2}e^{-(X_1 - X_2)}
			\right)
		}
		\smallskip \\
		+ c_{3} \displaystyle{ 
			\left(
			\frac{a_{1}\overline{a}_{2}\overline{b}_{1}b_{2}}{\overline{a}_{1}a_{2}b_{1}\overline{b}_2}e^{i\Theta_{12}}
			+
			\frac{\overline{a}_{1}a_{2}b_{1}\overline{b}_2}{a_{1}\overline{a}_{2}\overline{b}_{1}b_{2}}e^{-i\Theta_{12}}
			\right)	
		}
		\end{array}
		\right]^2}
},
\end{eqnarray}
where
\begin{eqnarray}
d_{jk}:=
\frac{(\alpha_{j}\overline{\beta}_{k} - \beta_{j}\overline{\alpha}_{k})(\lambda_{j} - \overline{\lambda}_{k})^3}{\lambda_{j}\overline{\lambda}_{k}},~
f_{jk}:=
\frac{(\alpha_{j}\beta_{k} - \beta_{j}\alpha_{k})(\lambda_{j} - \lambda_{k})^3}
{\lambda_{j}\lambda_{k}}.
\end{eqnarray}
For simplicity, we define
\begin{eqnarray}
\delta_{1}:=\frac{|a_{1}|^2}{|b_{1}|^2},~~
\delta_{2}:=\frac{|a_{2}|^2}{|b_{2}|^2},~~
\delta_{3}:=\frac{a_{1}\overline{a}_{2}}{b_{1}\overline{b}_{2}},~~
\delta_{4}:=\frac{a_{1}b_{2}}{a_{2}b_{1}},
\end{eqnarray}
which implies 
\begin{eqnarray}
|\delta_{3}|^2=\delta_{1}\delta_{2},~~ |\delta_{4}|^{2}=\frac{\delta_{1}}{\delta_{2}},~~
\frac{\delta_{3}}{|\delta_{3}|}=\frac{\delta_{4}}{|\delta_{4}|}.
\end{eqnarray}
Therefore, we can rewrite $\delta_{3}$ and $\delta_{4}$ in the polar form
\begin{eqnarray}
\delta_{3}=(\delta_{1}\delta_{2})^{\frac{1}{2}}e^{\frac{i\phi}{2}},~~
\delta_{4}=(\frac{\delta_{1}}{\delta_{2}})^{\frac{1}{2}}e^{\frac{i\phi}{2}}.
\end{eqnarray}

\section{Non-unitary Two-Soliton}
\label{Non-unitary 2-soliton}

Below are miscellaneous results on non-unitary two-soliton solutions with complex valued action density. 

In this case, we consider the unreduced non-unitary solution \eqref{J_nonunitary} in Theorem 3.4. The input data are not only $(\theta, \Lambda)$ in \eqref{2ss-matrix-pair} but also $(\rho, \Xi)$  given by 
\begin{eqnarray}
\rho=\left(
\begin{array}{cc}
a_{1}^2e^{\eta_{1}} & a_{2}^2e^{\eta_{2}}
\\
b_{1}^2e^{-\eta_{1}} & b_{2}^2e^{-\eta_{2}}
\end{array}
\right),~~
\Xi=
\left(
\begin{array}{cc}
\mu_{1} & 0
\\
0 & \mu_{2}
\end{array}
\right), 
\end{eqnarray}

where 
\begin{eqnarray}
	\label{eta_tilde}
	 \eta_{j} :=\mu_{j}\overline{\alpha}_{j}z + \overline{\beta}_{j}\widetilde{z} + \mu_{j}\overline{\beta}_{j}w + \overline{\alpha}_{j}\widetilde{w}, ~j=1, 2.
\end{eqnarray}
The NL$\sigma$M action density for the non-unitary $\hat{J}$ in  Sec. 4.3 is
\begin{eqnarray}
	{\cal{L}}_\sigma&=&
	-\frac{1}{16\pi}
	\mbox{Tr}\left[(\partial_{\mu}\hat{J})\hat{J}^{-1}(\partial^{\mu}\hat{J})\hat{J}^{-1}\right] \nonumber  \\
	&\!\!\!\! = \!\!\!\!&
	\label{NL Sigma term_2-Soliton_form 2_nonunitary}
	\frac{
		\left\{ 
		\begin{array}{l}
			~c_{1}c_{2}
			\displaystyle{\left[
				d_{11}\cosh^2\widetilde{X}_{2} + d_{22}\cosh^2\widetilde{X}_{1} 
				\right]
			}
			\medskip \\
			\!\!+ c_{1}c_{3}
			\left[
			\displaystyle{
				d_{12} \cosh^2\left( 
				\frac{\widetilde{X}_{1} + \widetilde{X}_{2} - i\widetilde{\Theta}_{12}}{2}
				\right)
				
				+ \widetilde{d}_{12}\cosh^2\left( 
				\frac{\widetilde{X}_{1} + \widetilde{X}_{2} + i\widetilde{\Theta}_{12}}{2}
				\right)
			}
			\right]
			\medskip \\
			\!\!- c_{2}c_{3}\left[
			\displaystyle{
				f_{12}~\!\sinh^2\left(
				\frac{\widetilde{X}_{1} - \widetilde{X}_{2} -i\widetilde{\Theta}_{12}}{2}
				\right)
				+ \widetilde{f}_{12}~\!\sinh^2\left(
				\frac{\widetilde{X}_{1} - \widetilde{X}_{2} +i\widetilde{\Theta}_{12}}{2}
				\right)
			}
			\right]
		\end{array}
		\!\!\!\right\}
	}
	{2\pi\left[
		c_{1}\cosh(
		\widetilde{X}_{1} + \widetilde{X}_{2} 
		)
		+ c_{2}\cosh(
		\widetilde{X}_{1} - \widetilde{X}_{2}
		)
		+ c_{3}\cos\widetilde{\Theta}_{12}
		\right]^2},~~~~~~
\end{eqnarray}
where
\begin{subequations}
\begin{eqnarray}
	c_{1}&\!\!\!\!:=\!\!\!\!&(\lambda_{1}-\lambda_{2})(\mu_{1}-\mu_{2}),~
	c_{2}:=(\lambda_{1}-\mu_{2})(\mu_{1}-\lambda_{2}),~
	c_{3}:=(\lambda_{1}-\mu_{1})(\lambda_{2}-\mu_{2}),
	\\
	d_{12}&\!\!\!\!:=\!\!\!\!&
	\frac{(\alpha_{1}\overline{\beta}_{2} - \beta_{1}\overline{\alpha}_{2})(\lambda_{1} - \mu_{2})^3}{\lambda_{1}\mu_{2}},~
	f_{12}:=
	\frac{(\alpha_{1}\beta_{2} - \beta_{1}\alpha_{2})(\lambda_{1} - \lambda_{2})^3}{\lambda_{1}\lambda_{2}}, 
	\\
	\widetilde{d}_{12}&\!\!\!\!:=\!\!\!\!&
	\frac{(\alpha_{1}\overline{\beta}_{2} - \beta_{1}\overline{\alpha}_{2})(\lambda_{1} - \mu_{2})^3}{\lambda_{1}\mu_{2}},~
	\widetilde{f}_{12}:=
	\frac{(\overline{\alpha}_{1}\overline{\beta}_{2} - \overline{\beta}_{1}\overline{\alpha}_{2})(\mu_{1} - \mu_{2})^3}{\mu_{1}\mu_{2}}, 
	\\
	\widetilde{X}_{j}&\!\!\!\!:=\!\!\!\!&\xi_{j} +  \eta_{j}  + \log \delta_{j}, ~\delta_{j}:=|a_{j}|^2 / |b_j|^2,
	\\
	\widetilde{\Theta}_{12}&\!\!\!\!:=\!\!\!\!&\Theta_{1} - \Theta_{2} + \phi, ~
	\Theta_{j}:=-i(\xi_{j}- \eta_{j}), ~j=1, 2,
	\\
	\phi&\!\!\!\!:=\!\!\!\!&2\mbox{Arg}(a_{1}\overline{a}_{2} / b_{1}\overline{b}_{2})
	=2\mbox{Arg}(a_{1}b_{2} / a_{2}b_{2}).
\end{eqnarray}
\end{subequations} 
\section{Data of the Double-Pole Solution }
\label{Double-pole data}
Substituting \eqref{Double-pole solution} into \eqref{Sylvester eq_hermitian}, we get the explicit form of $\Omega:=(\Omega_{ij})$, where
\begin{eqnarray}
	\label{Omega_elements}
	\begin{array}{l}
		\Omega_{11}
		=\displaystyle{\frac{-2}{(\lambda-\overline{\lambda})^3}}
		\left\{
		[(\lambda-\overline{\lambda})^2|\xi^{\prime}|^2 - 2] \mbox{cosh}(\xi+\overline{\xi})
		+ (\lambda - \overline{\lambda})(\xi^{\prime} - \overline{\xi}^{\prime})\mbox{sinh}(\xi+\overline{\xi})
		\right\},
		\smallskip \\
		\Omega_{12}
		=\displaystyle{\frac{-2}{(\lambda-\overline{\lambda})^2}}
		\left[
		(\lambda-\overline{\lambda})\overline{\xi}^{\prime}\mbox{sinh}(\xi+\overline{\xi})+\mbox{cosh}(\xi+\overline{\xi})
		\right],
		\smallskip \\
		\Omega_{21}
		=\displaystyle{\frac{-2}{(\lambda - \overline{\lambda})^2}}
		\left[
		(\lambda-\overline{\lambda})\xi^{\prime}\mbox{sinh}(\xi+\overline{\xi})-\mbox{cosh}(\xi+\overline{\xi})
		\right],
		\smallskip \\
		\Omega_{22}
		=\displaystyle{\frac{-2}{\lambda-\overline{\lambda}}}\mbox{cosh}(\xi+\overline{\xi}).
	\end{array}
\end{eqnarray}
The determinant of $\Omega$ is
\begin{eqnarray}
	\label{Omega_determinant}
	|\Omega|
	=\displaystyle{\frac{4}{(\lambda - \overline{\lambda})^4}}
	\left[
	(\lambda-\overline{\lambda})^2|\xi^{\prime}|^2 
	- \cosh^2(\xi + \overline{\xi})
	\right].
\end{eqnarray}
Substituting \eqref{Double-pole solution}, \eqref{Omega_elements}, and \eqref{Omega_determinant} into \eqref{J_unitary}, we get the explicit form of the $J$-matrix 
\begin{eqnarray}
	\hat{J}:=
	\displaystyle{\frac{1}{\Delta}}
	\left(
	\begin{array}{cc}
		\Delta_{11} & \Delta_{12} \\
		\Delta_{21} & \Delta_{22}
	\end{array}
	\right),  \nonumber 
\end{eqnarray}
where
\begin{eqnarray*}
	\label{J hat_doulbe pole solution_BDT_}
	\begin{array}{l}
		\Delta
		:=
		2\overline{\lambda}^2
		\left\{\displaystyle{\frac{1}{2}}\left[1+\mbox{cosh}(2(\xi+\overline{\xi}))\right] - (\lambda- \overline{\lambda})^2|\xi^{\prime}|^2
		\right\}
		\medskip \\
		\Delta_{11}
		:=
		\left\{
		\begin{array}{l}
			(\lambda - \overline{\lambda})^2
			\left[
			(\lambda \xi^{\prime}-\overline{\lambda}\overline{\xi}^{\prime})
			-2|\lambda|^2|\xi^{\prime}|^2
			\right]
			\smallskip \\
			+
			\displaystyle{\frac{(\lambda^2 + \overline{\lambda}^2)}{2}}
			\left[1+\mbox{cosh}(2(\xi+\overline{\xi}))
			\right]
			+
			\displaystyle{\frac{(\lambda^2 - \overline{\lambda}^2)}{2}}
			\mbox{sinh}(2(\xi+\overline{\xi}))
		\end{array}
		\right\},~
		\Delta_{22}=\overline{\Delta}_{11}, ~~
		\medskip \\
		\Delta_{12}
		:=
		(\lambda - \overline{\lambda})^2
		(\overline{\lambda} \overline{\xi}^{\prime}e^{2\xi}+\lambda \xi^{\prime}e^{-2\overline{\xi}})
		+
		(\lambda^2-\overline{\lambda}^2)\mbox{cosh}(\xi+\overline{\xi})e^{\xi-\overline{\xi}},~
		\Delta_{21}
		=-\overline{\Delta}_{12}.
	\end{array}
\end{eqnarray*}
\begin{eqnarray*}
	\partial_{z}\Delta
	&\!\!\!\!=\!\!\!\!& -2\overline{\lambda}^2
	\left\{
	(\lambda+\overline{\lambda})^2(\alpha\overline{L}^{\prime} + \overline{\alpha}L^{\prime})
	+(\lambda\alpha + \overline{\lambda}\overline{\alpha}) ~\!\mbox{sinh}\left[2(L+\overline{L})\right]
	\right\},~~~~~~~~~~~~~~~~~~~~~~~~
	\\
	\partial_{\widetilde{z}}\Delta
	&\!\!\!\!=\!\!\!\!&-2\lambda^2(\beta+\overline{\beta})~\!\mbox{sinh}[2(L+\overline{L})],
	\\
	\partial_{w}\Delta
	&\!\!\!\!=\!\!\!\!&\partial_{z}\Delta\big|_{\alpha \rightarrow \beta},~~
	\partial_{\widetilde{w}}\Delta = \partial_{\widetilde{z}}\Delta\big|_{\beta \rightarrow \alpha}.
\end{eqnarray*}
\begin{eqnarray*}
	\partial_{z}\Delta_{11}
	&\!\!\!\!=\!\!\!\!&
	\left\{
	\begin{array}{l}
		~~\!(\lambda+\overline{\lambda})^2
		\left[
		(\lambda\alpha-\overline{\lambda}\overline{\alpha})
		-2|\lambda|^2(\alpha\overline{L}^{\prime} + \overline{\alpha}L^{\prime})
		\right]
		\smallskip \\
		\!\!+(\lambda\alpha+\overline{\lambda}\overline{\alpha})
		\left[
		(\lambda^2 + \overline{\lambda}^2)~\!\mbox{sinh}[2(L + \overline{L})]
		+(\lambda^2 - \overline{\lambda}^2)~\!\mbox{cosh}[2(L + \overline{L})]
		\right]
	\end{array}
	\right\} ~~~~~~~~
	\smallskip \\
	\partial_{\widetilde{z}}\overline{\Delta}_{11}
	&\!\!\!\!=\!\!\!\!&
	(\beta + \overline{\beta})
	\left[
	(\lambda^2 + \overline{\lambda}^2)~\!\mbox{sinh}[2(L + \overline{L})]
	-(\lambda^2 - \overline{\lambda}^2)~\!\mbox{cosh}[2(L + \overline{L})]
	\right], ~~~~~~~~~~
	\smallskip \\
	\partial_{w}\Delta_{11}
	&\!\!\!\!=\!\!\!\!&\partial_{z}\Delta_{11}\big|_{\alpha \rightarrow \beta},~
	\partial_{\widetilde{w}}\overline{\Delta}_{11}
	=\partial_{\widetilde{z}}\overline{\Delta}_{11}\big|_{\beta \rightarrow \alpha}.
\end{eqnarray*}
\begin{eqnarray*}
	\partial_{z}\Delta_{12} &\!\!\!\!=\!\!\!\!&
	\left\{
	\begin{array}{l}
		~~\!(\lambda + \overline{\lambda})^2
		\left[
		(\overline{\lambda}\overline{\alpha} + 2|\lambda|^2\alpha\overline{L}^{\prime})e^{2L}
		+
		(\lambda\alpha - 2|\lambda|^2\overline{\alpha}L^{\prime})e^{-2\overline{L}}
		\right]
		\smallskip \\
		\!\!+(\lambda^2 - \overline{\lambda}^2)
		\left[
		(\lambda\alpha + \overline{\lambda}\overline{\alpha})~\!\mbox{sinh}[L+\overline{L}]
		+
		(\lambda\alpha - \overline{\lambda}\overline{\alpha})~\!\mbox{cosh}[L+\overline{L}]
		\right]e^{L-\overline{L}}
	\end{array}
	\right\}, ~~~~~~~
	\smallskip \\
	\partial_{\widetilde{z}}\overline{\Delta}_{12}
	&\!\!\!\!=\!\!\!\!&
	\left\{
	\begin{array}{l}
		~~\!2(\lambda+\overline{\lambda})^2
		\left(
		\lambda\overline{\beta}L^{\prime}e^{2\overline{L}}
		-\overline{\lambda}\beta\overline{L}^{\prime}e^{-2L}
		\right)
		\smallskip \\
		\!\!-(\lambda^2 - \overline{\lambda}^2)\left[
		(\beta+\overline{\beta})~\!\mbox{sinh}(L+\overline{L})
		-(\beta-\overline{\beta})~\!\mbox{cosh}(L+\overline{L})
		\right]e^{-(L-\overline{L})}
	\end{array}
	\right\},
	\smallskip \\
	\partial_{w}\Delta_{12}
	&\!\!\!\!=\!\!\!\!&
	\partial_{z}\Delta_{12}\big|_{\alpha \rightarrow \beta},~
	\partial_{\widetilde{w}}\overline{\Delta}_{12}=\partial_{\widetilde{z}}\overline{\Delta}_{12}\big|_{\beta \rightarrow \alpha}.
\end{eqnarray*}

\end{appendices}



\end{document}